\documentclass[12pt,a4paper]{article}
\usepackage{epsfig}

\newcommand{\nc}{\newcommand}
\nc{\beq}{\begin{equation}} \nc{\eeq}{\end{equation}}
\nc{\beqa}{\begin{eqnarray}} \nc{\eeqa}{\end{eqnarray}}

\begin{document}
\begin{flushright}AEI-2010-163
\end{flushright}
\begin{center}

\vspace{1cm}

{\bf \large ON FORM FACTORS IN $\mathcal{N}=4$ SYM} \vspace{2cm}

{\bf \large L. V. Bork$^{2}$, D. I. Kazakov$^{1,2}$, G. S.
Vartanov$^{3}$}\vspace{0.5cm}

{\it $^1$Bogoliubov Laboratory of Theoretical Physics, Joint
Institute for Nuclear Research, Dubna, Russia, \\
$^2$Institute for Theoretical and Experimental Physics, Moscow, Russia, \\
$^3$Max-Planck-Institut f\"ur Gravitationsphysik,
Albert-Einstein-Institut 14476 Golm, Germany.}\vspace{1cm}

\abstract{In this paper we study the form factors for the half-BPS
operators $\mathcal{O}^{(n)}_I$ and  the $\mathcal{N}=4$ stress
tensor supermultiplet current $T^{AB}$ up to the second order of
perturbation theory and for the Konishi operator $\mathcal{K}$ at
first order of perturbation theory in the $\mathcal{N}=4$ SYM theory
at weak coupling. For all the objects we observe the exponentiation
of the IR divergences with two anomalous dimensions: the cusp
anomalous dimension and the collinear anomalous dimension. For the
IR finite parts we obtain a similar situation as for the gluon
scattering amplitudes, namely, apart from the case of $T^{AB}$ and
$\mathcal{K}$ the finite part has some remainder function which we
calculate up to the second order. It involves the generalized
Goncharov polylogarithms of several variables. All the answers are
expressed in terms of  the integrals  related to the dual conformal
invariant ones which might be a signal of integrable structure
standing behind the form factors.}
\end{center}



\newpage

\tableofcontents{}\vspace{0.5cm}

{\bf References} \hfill {\bf 31}

\renewcommand{\theequation}{\thesection.\arabic{equation}}

\section{Introduction}\label{s1}

Much attention in the past few years has been dedicated to the study
of the planar limit for the scattering amplitudes in the
$\mathcal{N}=4$ SYM theory. It is believed that the hidden
symmetries responsible for integrability properties of
$\mathcal{N}=4$ SYM completely fix the structure of the amplitudes
(the $S$-matrix of the theory)
\cite{BeisertYangianRev,BeisertYangianAmpl}. One of the possible
views on this subject is that the answers for the amplitudes are
expressed in terms of the scalar integrals which are
pseudo-conformal invariant in momentum space \cite{dualKorch} which
appear in the unitarity-based calculation of the scattering
amplitudes pioneered in papers \cite{Bern1}.

The (dual)conformal symmetry at weak coupling regime can be extended
to the $\mathcal{N}=4$ supersymmetric version and can be fused with the
original $\mathcal{N}=4$ superconformal symmetry  to the so-called
Yangian symmetry \cite{yangian} which is governed by Yangian
infinite dimensional algebra. The Yangian like symmetries are common
features of the integrable systems \cite{BeisertYangianRev}.

At strong coupling the computation of  the amplitudes in $\mathcal{N}=4$
SYM can be reduced via $AdS/CFT$ to the computation of the open string
scattering amplitudes in $AdS_5$, with strings ending on $D3$-brane
positioned at some fixed value of the radial $AdS_5$ coordinate $z$, in the
quasi-classical regime \cite{minSurface4point} which in turn can be
formulated as the problem of finding the minimal surface in $AdS_5$ with
special boundary condition (see \cite{Alday:2008yw} for review).
This problem has  recently  been reduced to that of solving the set
of functional equations for the conformal invariant cross ratios as
functions of the spectral parameters-- the so-called $Y$-system
\cite{Alday:2010vh}. The $Y$-systems usually appear in integrable
systems \cite{Y-review} which is another hint that  the $\mathcal{N}=4$
amplitudes have some underling integrable structure.

In strong coupling regime the natural generalization of the $Y$-system
for the amplitudes is the $Y$-system for the form factors
\cite{Maldacena:2010kp}: the matrix elements of the form \beq
\langle 0|\mathcal{O}|p_1^{\lambda_1} \ldots p_n^{\lambda_n}\rangle.
\eeq where $\mathcal{O}$ is some gauge invariant operator which acts
on vacuum and produces some state $|p_1^{\lambda_1} \ldots
p_n^{\lambda_n} \rangle$ with momenta $p_1 \ldots p_n$ and
helicities $\lambda_1 \ldots \lambda_n$. In the dual string theory
this matrix element can be described via the open string scattering
amplitudes, with strings ending on $D3$-brane positioned at some
fixed value of the radial $AdS_5$ coordinate $z$, in the presence of some
closed string state \cite{Maldacena:2010kp}.

One can wonder whether these objects at weak coupling possess similar
features as the amplitudes or in other words whether form factors are
influenced by the Yangian symmetry (or some analog of it)  and whether
they are fixed by it. Also the general structure of the form factors
at weak coupling should be understood.

Being inspired by the two-loop calculation of the form factor
associated with the operator $\mathcal{V}_X$ from the stress-tensor
superconformal multiplet of the $\mathcal{N}=4$ SYM theory performed long ago
by van Neerven \cite{vanNeerven:1985ja} we would like to study
systematically some types of form factors in planar $\mathcal{N}=4$
SYM at weak coupling for half-BPS operators $ \mathcal{O}^{(n)}_I$
and the Konishi operator $\mathcal{K}$. For the former type of
operators there was recently  an interest in studying the
correlation functions and their connection to the amplitudes and the
Wilson loops \cite{Eden:2010zz}. A new kind of relation
has been proposed between the  logarithm for such
correlation functions and the double logarithm of the MHV gluon
scattering amplitudes.

The dual-conformal symmetry plays an important role in another
remarkable property of the $\mathcal{N}=4$ SYM -- the Wilson
loop/Amplitudes duality \cite{dual,Gorsky:2009ew}. In this duality
(we restrict ourselves to the most studied case of the MHV
amplitude\footnote{MHV (maximally helicity violating) amplitudes by
definition are called the amplitudes with all particles being
treated as outgoing and the net helicity $\lambda_{\Sigma}$ being
equal to $n-4$ where $n$ is the number of particles.} sector) the
dual-conformal symmetry is understood as conformal symmetry of
light-like Wilson loop constructed of the segments which satisfy the
following property:
\beq
x_{i,i+1}^{\mu}=x_i^{\mu}-x_{i+1}^{\mu}=p^{\mu}_i, \label{WLsegments}
\eeq
where $p^{\mu}_i$ are external momenta of the dual MHV amplitude.
The dual-conformal symmetry is broken on-shell for the amplitudes due to
the presence of the IR divergences  (these IR divergences
correspond to the UV divergences for dual Wilson loops); however,
the violation of the dual-conformal symmetry is controlled by the 1-loop
exact anomaly, which in turn can be used to make constraints for
the finite part of the corresponding MHV amplitude, i.e. one can
write the anomalous Ward identities allowing one to constraint the
finite parts (see, for example, review \cite{Alday:2008yw} for
details):
\begin{eqnarray}
\sum_{i=1}^{n}(2x_i^{\nu}x_i \partial_i -x_i^2 \partial_i^{\nu}) Fin
[\log {\cal W}_n] = \Gamma_{cusp} \sum_{i=1}^n \log
\frac{x_{i,i+2}^2}{x_{i-1,i+1}^2} x_{i,i+1}^{\nu},
\end{eqnarray}
where $\Gamma_{cusp}$ is the so-called cusp anomalous dimension
\cite{colK} known from a solution of the integral
equation \cite{Transc}. These identities can fix the finite part  for a
small number of legs/cusps (namely, one can fix it for Wilson loops
with $n=4$ and $n=5$ cusps) \cite{Alday:2008yw}.
In fact, the famous BDS conjecture was the simplest possible ansatz of
these identities, which is not precisely correct for a number of
external legs $\geq5$.

One may wonder if there is a similar duality for the form factors/Wilson
loops and one can use similar arguments to obtain
information on the finite parts of the form factors. We hope that our
calculation shades some light on the possibility of such duality.

The paper is organized as follows. In Sect. \ref{s2}, we present the
general considerations of the form factors  in the $\mathcal{N}=4$ SYM
theory and introduce the operators to be discussed later. In Sect.
\ref{s3}, we study the form factors for  both protected and
non-protected operators with naive conformal dimension $\Delta_0=2$
and confirm the results obtained long time ago in
\cite{vanNeerven:1985ja}. In Sect. \ref{s4}, we study the half-BPS
operators $\mathcal{O}^{(n)}_I$ for arbitrary conformal dimension
$\Delta_0=n$ and present the one- and two-loop calculations of the
corresponding form factors which suggest the exponentiation of the
IR divergences. Also, in the same section, we discuss the collinear
limit for which the finite parts take a simple form. In Sect.
\ref{s5}, we discuss in more detail the  dual conformal invariance of
the integrals contributing to the calculation of the form factors.
We conclude with some remarks concerning the form factors and Wilson
loop duality. In the appendices we give the  details of our
calculations. Appendix. A contains the Lagrangian of the $\mathcal{N}=4$
SYM theory together with the Feynman rules. In App. B we present the
analytic expressions for the integrals entering into our calculations
with their $\epsilon$-expansion. The results of our work have been
reported at the international conference devoted to the memory of
A.N. Vasiliev held in Saint-Petersburg on 18-21 October 2010.

\section{General considerations}\label{s2}

\subsection{Form factors in $\mathcal{N}=4$ SYM}\label{sb21}

Consider the Lagrangian
$\mathcal{L}_{\mathcal{N}=4}(\mathcal{W})$ for the $\mathcal{N}=4$ SYM theory
coupled to some external classical current $J$ through some gauge
invariant local operator $\mathcal{O}[\mathcal{W}]$
 (for all the details
concerning the explicit expression for the Lagrangian together with
Feynman rules we refer to App. A)
\beq\mathcal{L}_{\mathcal{N}=4}(\mathcal{W}) \to
\mathcal{L}_{\mathcal{N}=4}(\mathcal{W}) +
\mathcal{O}[\mathcal{W}]J, \eeq where we collectively refer to the
whole $\mathcal{N}=4$ on-shell multiplet as $\mathcal{W}$ which
consists of the physical gluon $A^{\mu}$ states with positive and
negative helicities, four gauginos $\lambda^{N}_{\alpha}$ with
positive and negative helicities and also six real scalar states
$\phi_{NM}$, where $N$ and $M$ are the $SU(4)_R$ indices, which can
also  be re-arranged into 3 complex  $\phi^I$ scalars, $I$ is
 an index of $SU(3)$ subgroup of $SU(4)_R$. By default we assume everywhere
the planar limit.

\begin{figure}[ht]
 \begin{center}
 \leavevmode
  \epsfxsize=5cm
 \epsffile{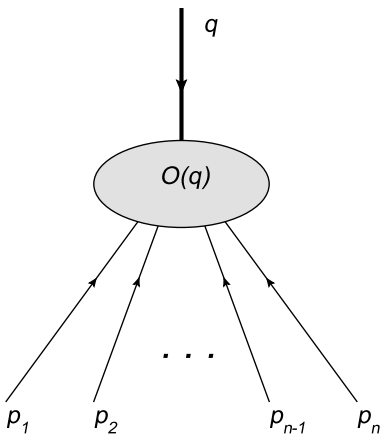}
 \end{center}\vspace{-0.2cm}
 \caption{Feynman diagram for the matrix element of the operator $\mathcal{O}$}\label{oper}
 \end{figure}

Then one can study the following processes where the operator
$\mathcal{O}$ acts on the vacuum and produces some state
$|p_1^{\lambda_1} \ldots p_n^{\lambda_n} \rangle$ with momenta $p_1
\ldots p_n$ and helicities $\lambda_1 \ldots \lambda_n$ \beq \langle
0|\mathcal{O}|p_1^{\lambda_1} \ldots p_n^{\lambda_n}\rangle. \eeq
Schematically, it is shown in Fig.\ref{oper}

This is a general situation in QFT and one can keep in
mind, for example, $\gamma^{*}\to \mbox{Jet's}$ process
\cite{Kazakov:1987jk} where we take into account all orders in
$\alpha_s$ but the first order in $\alpha_{em}$. In
perturbation theory the latter type of processes can be thought of as the
matrix elements of the following form:
 \beq \langle
0|j^{QCD}_{em}|p_1^{\lambda_1} \ldots p_n^{\lambda_n}\rangle, \eeq
where $j^{QCD}_{em}$ is the QCD quark electromagnetic current.

The matrix element $\langle 0|\mathcal{O}|p_1^{\lambda_1} \ldots
p_n^{\lambda_n}\rangle$ in some sense can also be viewed as the
generalization of the scattering amplitudes, which in ''all ingoing"
notation can schematically be written as $\langle 0|p_1^{\lambda_1}
\ldots p_n^{\lambda_n}\rangle$.

In the language of the dual string theory, in the $\mathcal{N}=4$ SYM
case this process can be described as an insertion of some close
string state (which corresponds to $\mathcal{O}$ local operator) on
the worldsheet in addition to $n$ open string states (which
corresponds to $|p_1^{\lambda_1} \ldots p_n^{\lambda_n}\rangle$
state in the dual theory).

For the construction of particular examples of the objects discussed
above we choose the following set of the gauge invariant operators
(we use the component notation of  the $\mathcal{N}=4$ SYM),  the
lowest stress tensor supermultiplet members: \beqa \mathcal{C}_{IJ}
&=& \mbox{Tr}(\phi_I\phi_J), \nonumber \\ \mathcal{V}_{I}^{J} &=&
\mbox{Tr}(\bar{\phi}^{J}\phi_I),\eeqa with naive mass dimension
$\Delta_0=2$ which coincides with the conformal dimension due to the
lack of quantum corrections. These operators can be viewed as the
lowest members of the stress-tensor  multiplet \beq
T^{AB}=\mbox{Tr}\left(W^AW^B - \frac 16 \delta^{AB}W^CW_C\right),
\eeq where $A,B, \ldots = 1, \ldots, 6$ are the $SO(6)_R \simeq
SU(4)_R$ indices, $I,J, \ldots =1,2,3 $ are the indices of $SU(3)$
subgroup of $SU(4)_R$, and $W^{A}$ is some constrained chiral
superfield in $\mathcal{N}=4$ superspace containing all components
of the $\mathcal{N}=4$ supermultiplet.

Other objects are the so-called half-BPS operators \beq
\mathcal{O}_I^{(n)}=\mbox{Tr}(\phi_I^n),\eeq whose naive mass
dimension coincides with conformal dimension $\Delta_0=n$ being
protected from the quantum corrections, and the lowest component of
the Konishi supermultiplet \beq
\mathcal{K}=\sum_{I=1}^{3}\mbox{Tr}(\bar{\phi}^I\phi_I), \eeq with
naive mass dimension $\Delta_0=2$ and has nonvanishing
anomalous dimension due to the presence of the UV divergences. The
calculation of this anomalous dimension has been intensively
discussed  during the last few years
\cite{Kotikov:2003fb,4loopKonsihi}.

Since the Konishi operator is not protected, the corresponding form
factors a priori do contain the UV divergences and hence must be UV
renormalized. It means that one has to consider the renormalized
form factor \beq \langle 0|\mathcal{K}_{R}|p_1^{\lambda_1} \ldots
p_n^{\lambda_n}\rangle, \eeq where \beq
\mathcal{K}_{R}=Z^{-1}_{K}\mathcal{K}_{B}. \eeq Here $Z_{K}$ is the
renormalization constant which appears due to the UV divergences and
which should be calculated to the same order of perturbation theory
as  the form factors. After such UV renormalization we are left only
with the IR divergences. All the statements  concerning the
Konishi operator are valid for the renormalized one.

We choose for simplicity the state $| p_1^{\lambda_1} \ldots
p_n^{\lambda_n}\rangle$ produced by the operator $\mathcal{O}$ to
consist of scalars only and then we can write it as $|p_1 \ldots
p_n\rangle$ without helicities. We also restrict
ourselves to the  states with the number of particles equal to
the naive mass dimension of the operator $\mathcal{O}$, i.e. we
consider the  states consisting of $\Delta_0$ scalars.

\subsection{Calculation  strategy}

For the calculation it is convenient to use the $\mathcal{N}=1$ formulation
of  $\mathcal{N}=4$ SYM and perform an explicit computation in terms of the
$\mathcal{N}=1$ superfields in momentum space.
 Our computation is
familiar,  from a diagrammatic point of view, to
perturbative computations of anomalous dimensions
\cite{4loopKonsihi}. However, there is a significant difference: each
of our supergraphs is UV finite except for  one.  So all the divergences that appear throughout the calculation
have the IR nature.

The operators $\mathcal{O}=\{ \mathcal{C}_{IJ}, \mathcal{V}_{I}^{J},
\mathcal{K}, \mathcal{O}_I^{(n)} \}$ can be considered as the lowest
components of the following $\mathcal{N}=1$
local operators:  \beqa \label{operators}
\mathcal{C}_{IJ} &=& \mbox{Tr}(\Phi_I\Phi_J),I \neq J \nonumber \\
\mathcal{V}_{I}^{J} &=&
\mbox{Tr}(e^{-gV}\bar{\Phi}^Je^{gV}\Phi_I), I \neq J  \nonumber \\
\mathcal{O}_I^{(n)} &=& \mbox{Tr}\left(\Phi_I^n\right),  \nonumber \\
\mathcal{K} &=& \sum_{I}\mbox{Tr}(e^{-gV}\bar{\Phi}^Ie^{gV}\Phi_I),
\eeqa where $\Phi_I$ are chiral $\mathcal{N}=1$ superfields, and $V$
is $\mathcal{N}=1$ real vector superfield (see App. A for details). The operators
$\mathcal{C}_{IJ},~ \mathcal{O}_I^{(n)}$ are chiral and
$\mathcal{V}_{I}^{J},~\mathcal{K}$ are non-chiral from the
$\mathcal{N}=1$ supersymmetric point of view.

We use the following notation for the form factor of the
corresponding operator \beq \mathcal{F}(p_1 \ldots p_n)=\langle p_1
\ldots p_n|\mathcal{O}(q)|0\rangle. \eeq

We expect the following factorization property for $\mathcal{F}$ to
hold:
 \beq \mathcal{F}(p_1 \ldots
p_n)=\mathcal{F}_{tree}(p_1 \ldots p_n)(1 + \mbox{loops}), \eeq
where $\mathcal{F}_{tree}$  stands for the tree level contribution,  and ''$\mbox{loops}$"
schematically denote the contributions of the next orders of PT. It is convenient to consider the ratio
$$
\mathcal{M}=\frac{\mathcal{F}}{\mathcal{F}_{tree}}=(1+loops)=\sum_{l=0}\lambda^l\mathcal{M}^{(l)},
$$
where $\lambda\equiv g^2 N_c$ is the  't Hooft coupling which stays fixed when $N_c\to \infty$.

Consider first the chiral case. To calculate the form factor  it is
convenient to consider the generating functional for the
one-particle irreducible super diagrams $\Gamma[\Phi^{cl},J]$ in
$\mathcal{N}=1$ superspace. It can be obtained from the generating
functional
$$
Z[j,J]=\int\mathcal{D}(\Phi_I,V, \ldots)\exp[S^{\mathcal{N}=4}+\int
d^6zJ(z) \mathcal{O}(z) + \int d^6z\mbox{Tr}(j(z)\Phi(z))],
$$
after Legendre transformation with respect to  external chiral
sources $j$ (note that the source $J$  is  untouched).  After
performing   the D-algebra  each supergraph gives  a local
contribution in $\theta$'s, and $\Gamma[\Phi^{cl},J]$  can be written
as ( we imply the mass shell condition $p_i^2=0$ when performing the
D-algebra)  \beqa &&
\Gamma[\Phi^{cl},J]=\sum_{l=0}\lambda^l\Gamma^{(l)}[\Phi^{cl},J]
\\ \nonumber && \makebox[-2em]{} = \sum_{l=0}\lambda^l\!\int d^4p_1 \ldots d^4p_n~d^6z
~J(-q,\theta)\mbox{Tr} \left( \Phi^{cl}(-p_1,\theta)
\ldots\Phi^{cl}(-p_n,\theta) \right) \mathcal{M}^{(l)}(p_1, \ldots
p_n)\!+\!O(J^2), \eeqa where $d^6z = d^4q d^2\theta$,
$\mathcal{M}^{(l)}$ is given by the sum of scalar integrals. Thus,
\beqa
\mathcal{M}^{(l)}(p_1...p_n)=\frac{\delta^{n+1}\Gamma^{(l)}}{\delta\Phi^{cl}...\delta\Phi^{cl}
\delta J}\Big|_{p_i^2=0,~\theta=0, \Phi^{cl}=0,J=0}. \eeqa We stress
that on-shell condition $p_i^2=0$ and momenta conservation $q+p_1+
\ldots+p_n=0$ are implemented to obtain the latter expression.

The situation is a bit more involved in the nonchiral case. All the
integrals in $\Gamma[\Phi^{cl},\bar\Phi^{cl}, \mathcal{J}]$
($\mathcal{J}$ is  a non-chiral source) are now in full
$\mathcal{N}=1$ superspace $\int d^8z$, where $~d^8z=d^4qd^4\theta$
and the expression for $\Gamma[\Phi^{cl},\bar\Phi^{cl},
\mathcal{J}]$ contains extra terms \beqa &&
\Gamma[\Phi^{cl},\bar\Phi^{cl}, \mathcal{J}] =
\sum_{l=0}\lambda^l\Gamma^{(l)}[ \Phi^{cl},\bar\Phi^{cl},
\mathcal{J}] = \label{nc}
\\ \nonumber && \makebox[-1em]{} = \sum_{l=0}\lambda^l\int d^4p_1 \ldots d^4p_n~d^8z
~\mathcal{J}(-q,\theta,\bar\theta)\left[ \mbox{Tr} \left(
\bar\Phi^{cl}(-p_1,\bar\theta) \ldots\Phi^{cl}(-p_n,\theta)
\right) \mathcal{M}^{(l)}(p_1, \ldots p_n) \right. \nonumber \\
&&\hspace{1cm}  \left. +\mbox{Tr} \left( \bar
D^{\dot\beta}\bar\Phi^{cl}(-p_1,\bar\theta) \ldots D^\alpha
\Phi^{cl}(-p_n,\theta) \right) \mathcal{M}_{\dot\beta
\alpha}^{(l)}(p_1, \ldots p_n)\right. \nonumber \\&&\hspace{1cm}
\left. + \mbox{Tr} \left( \bar D^2\bar\Phi^{cl}(-p_1,\bar\theta)
\ldots D^2\Phi^{cl}(-p_n,\theta) \right) \mathcal{M}_2^{(l)}(p_1,
\ldots p_n)\right] + O(\mathcal{J}^2)\nonumber . \eeqa

From the point of view of $\mathcal{N}=1$ superspace the additional
terms correspond to the operators of higher dimension and one
actually has a mixing of several operators. However, from the point
of view of components, one can always consider a projection on a
particular component of a superfield and we choose the scalar
component insofar. Then, the last terms of eq. (\ref{nc}) are
irrelevant  for our calculation and can be dropped.

We perform all the calculations  in the formalism of $\mathcal{N}=1$
superspace and at the end take the projection to $\theta=\bar
\theta=0$. There are pluses and minuses of this approach. The big
advantage is the drastic reduction of the number of diagrams
compared to the component case together with the simplified form of
the scalar integrals. Its disadvantage is that  we do not use the
power of the on-shell $\mathcal{N}=4$ covariant methods used in
perturbative studies of the amplitudes \cite{N=4onshellmethods} (see
also recent \cite{He:2010ju}). The application of this method for
the calculation of the form factors when some legs are off-shell
requires some modification.

\subsection{IR finite observables based on form factors}\label{sb22}

As the amplitudes, the form factors $\langle
0|\mathcal{O}|p_1^{\lambda_1} \ldots p_n^{\lambda_n}\rangle$ with
on-shell momenta are, strictly speaking, ill-defined in
$D=4$-dimensional  space-time due to the presence of the IR
divergences, and, hence, some IR regulator must be introduced -- in
our case it is the parameter $\mu$ coming from the dimensional
regularization which also breaks the conformal symmetry. In other
words,  one may say that $\langle 0|\mathcal{O}|p_1^{\lambda_1}
\ldots p_n^{\lambda_n} \rangle$ are the intermediate objects, and
the true physical quantities are the IR safe observables constructed
of $\langle 0|\mathcal{O}|p_1^{\lambda_1} \ldots
p_n^{\lambda_n}\rangle$ and which are free from the IR regulator
(see, for example, the discussion of the IR finite observables for
$\mathcal{N}=4$ SYM and $\mathcal{N}=8$ SUGRA in
\cite{finite1,finite2}). Indeed, as in QCD for $\gamma^{*} \to
\mbox{Jet's}$ processes we are really interested in the total cross
section $\sigma_{tot}(\gamma^{*} \to \mbox{Jet's})$ or some
differential distributions rather than in the matrix elements
$\langle 0|j^{QCD}_{em}|p_1^{\lambda_1} \ldots
p_n^{\lambda_n}\rangle$ themselves. This kind of observables are IR
finite due to the \textit{Kinoshita-Lee-Nauenberg (KLN) theorem}
which states that it is not sufficient to consider only the
processes with the fixed number of final particles. To get the
physical result, one has to include all the processes allowed by
conservation laws  in the same order of perturbation theory with
emission of extra soft quanta and integrate over their momenta.
Practically, if the dimensional regularization is used (the IR
divergences manifest themselves through the appearance of the
$1/\epsilon$ poles), part of the poles cancel between the loop
integrals from the virtual contributions and the phase space
integrals from
 the real
contributions coming from the processes with additional particles,
 while others are absorbed
in the functions describing probability distributions of the initial
and final states (in \cite{finite1,finite2}  we call them
initial-collinear and final-collinear divergences which appear as a
collinear configuration of initial and final particles).

Consider, for instance,  the total cross section $\sigma_{tot}$ for
the process $$J \to \mbox{anything from $\mathcal{N}=4$
supermultiplet}$$ for classical current $J$ coupled to
$\mathcal{N}=4$ through some local gauge invariant operator
$\mathcal{O}$.  Due to the optical theorem \beq \sigma_{tot}(s) \sim
\frac 1sIm_{s} \left[ \int d^Dx
\exp(-iqx)\langle\mathcal{O}(x)\mathcal{O}(0)\rangle
\right],~q^2=-s, \eeq The two-point function for the operators
$\mathcal{O}$  apart from the canonical mass dimension $\Delta_0$
can have anomalous dimension $\gamma=\gamma(\lambda)$ being a
function of the coupling constant \beq
\langle\mathcal{O}(x)\mathcal{O}(0)\rangle\sim\frac{1}{(x^2)^{(\Delta_0(1-\epsilon)+\gamma)}},
\eeq After some calculation this gives  the total cross-section \beq
\sigma_{tot}(s) \sim \frac{1}{\Gamma(\Delta_0 + \gamma)
\Gamma(\Delta_0 + \gamma - 1)} \frac{1}{s^{3 - \Delta_0 - \gamma}},
\eeq and its asymptotic at weak and strong couplings can be studied (compare this with C.3 from \cite{hofmal}).

In $\mathcal{N}=4$ SYM, as in any conformal theory, the operator
$\mathcal{O}$ is protected, which means that it does not receive
quantum corrections and $\gamma=0$.  Then the cross section is
independent of the coupling constant  and behaves like $\sim
C/s^{3-\Delta_0}$. From the latter expression it might seem that we
get violation of unitarity since we can get increasing cross
sections for protected operators with conformal dimension greater
than $3$. But it is not the case since the statement about the
unitarity holds only for the operators which give rise to renormalizable interactions, i.e.
with conformal dimension less than $3$.

If one is interested not in $\sigma_{tot}$ but in some differential
distributions,  then the optical theorem is not very useful any more, and
direct computations must be done. The form factors discussed here can
be viewed as the building blocks in the same sense as the amplitudes for the
inclusive cross sections.

\section{Form factors with $\Delta_0=2$}\label{s3}

In the following two sections we give explicit results for the
direct diagrammatic computation of the form factors of the
operators introduced above in the planar limit.\footnote{$g
\rightarrow 0$ and $N_c \rightarrow \infty$ so that
$\lambda=g^2N_c=$fixed.} More concretely we present the results for
non-chiral operators $\mathcal{V}_{I}^{J},~\mathcal{K}$  in the
leading order in  $\lambda$   and for the chiral operators
$\mathcal{C}_{IJ},~\mathcal{O}_I^{(n)}$ in the next-to-leading
order. The dimensional regularization (dimensional reduction to be
precise) with $D \ = \ 4 - 2 \epsilon$ is used. All the divergences
except for the specially mentioned cases have the IR (both the
infrared itself and collinear) nature. The Feynman rules for the
supergraphs are given in App. A. The complete list of all the
necessary scalar integrals is given in App. B.

\subsection{$\mathcal{C}_{IJ},~\mathcal{V}_{I}^{J}$ and $~\mathcal{K}$
form factors at 1-loop}\label{sb31}

\begin{figure}[ht]\vspace{0.3cm}
 \begin{center}
 \leavevmode
  \epsfxsize=15cm
 \epsffile{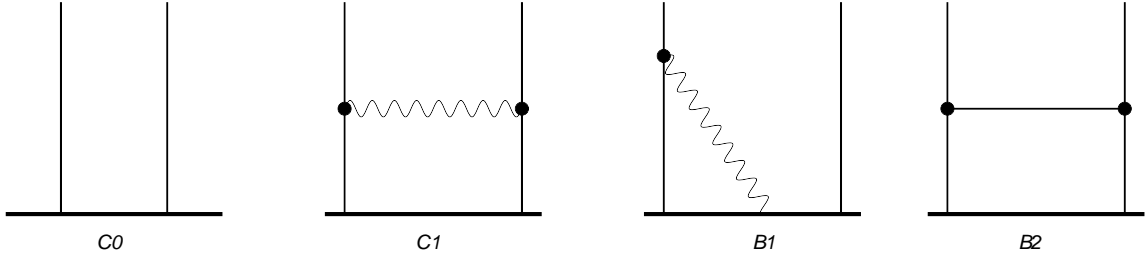}
 \end{center}\vspace{-0.2cm}
\caption{The relevant supergraphs. The internal black lines
correspond to chiral propagators $\langle \bar{\Phi}_I^a\Phi_J^b
\rangle$, wavy lines correspond to vector $\langle V^aV^b \rangle$
propagator (see App. A). $C0$ is the tree level diagram, and the
rest are one-loop ones. External lines are $\Phi$ or $\bar{\Phi}$,
and the lower bold line represents the insertion of the
corresponding operator in modern notation. For the chiral operator
$\mathcal{C}_{IJ}$ only the  diagrams $C0$ and $C1$ contribute,
while for non-chiral operators $\mathcal{V}_{I}^{J}$ and
$~\mathcal{K}$  the other two ($B1$ and $B2$) are also relevant.
}\label{supgr1}
 \end{figure}

The corresponding tree level and one-loop Feynman diagrams are shown in Fig.\ref{supgr1}.

For the chiral operator $\mathcal{C}_{IJ}$,  after performing the
$D$-algebra for the supergraph $C1$, the resulting expression is \beq
C1=\mbox{Tr} \left(
\Phi^{cl}_{I}\Phi^{cl}_{J}\right)~2s_{12}~G_1(s_{12}), \eeq where
$s_{ij}=(p_i+p_j)^2$, so that \beq
\mathcal{M}^{(1)}=2s_{12}~G_1(s_{12}), \eeq where the scalar
integral $G_1(s_{12})$ is given in App. B Hereafter we will suppress
index $cl$ in $\Phi$ and $\bar{\Phi}$.

For the non-chiral operator $\mathcal{V}_{I}^{J}$ after performing
the $D$-algebra one has \beqa C1
&=&2\left((-G_0(p_1^2)-G_0(p_2^2)+G_0(s_{12})
+(s_{12}-p_1^2-p_2^2)G_1(s_{12}))~\mbox{Tr} \left(
\bar{\Phi}^J\Phi_I \right) \right. \nonumber \\&&\left.
  -G_1^{\alpha\dot{\beta}}(s_{12})~\mbox{Tr} \left(
  \bar{D}^{\dot{\beta}}\bar{\Phi}^JD^{\alpha}\Phi_I \right)
 +G_1(s_{12})~\mbox{Tr} \left( \bar{D}^{2}\bar{\Phi}^JD^{2}\Phi_I \right) \right)
 ,\nonumber\\
B1&=& 2G_0(p_i^2)~\mbox{Tr} \left( \bar{\Phi}^J\Phi_I \right) ,\\
B2&=&2\left(\!-G_0(s_{12})\mbox{Tr} \left( \bar{\Phi}^J\Phi_I
\right) \!+\! G_1^{\alpha\dot{\beta}}(s_{12}) \mbox{Tr} \left(
\bar{D}^{\dot{\beta}}\bar{\Phi}^JD^{\alpha}\Phi_I \right) \!+\!
G_1(s_{12})\mbox{Tr} \left( \bar{D}^{2}\bar{\Phi}^JD^{2}\Phi_I
\right)\! \right) ,\nonumber \eeqa where all the scalar integrals
are also given in App.B. So keeping only the terms that are relevant
for our discussion, which are  proportional to
$\mbox{Tr}\left(\bar{\Phi}^{J}\Phi_{I}\right)$,  one gets \beq
\mathcal{M}^{(1)}=C1(s_{12},p_1^2,p_2^2)+B1(p_1^2)+B1(p_2^2)+B2(s_{12})=2(s_{12}-p_1^2-p_2^2)
G_1(s_{12}),\eeq The integral $G_1$ is UV finite which reflects the
fact that $\mathcal{V}_{I}^{J}$ is a protected operator.

For the non-chiral Konishi operator $\mathcal{K}$, after performing
the $D$-algebra  one has \beqa C1
&=&6\left((-G_0(p_1^2)-G_0(p_2^2)+G_0(s_{12})
+(s_{12}-p_1^2-p_2^2)G_1(s_{12}))\sum_I^3\mbox{Tr} \left(
\bar{\Phi}^I\Phi_I \right) \right. \nonumber \\&&\left.
-G_1^{\alpha\dot{\beta}}(s_{12})\sum_I^3\mbox{Tr} \left(
\bar{D}^{\dot{\beta}}\bar{\Phi}^ID^{\alpha}\Phi_I \right)
+G_1(s_{12})\sum_I^3\mbox{Tr} \left(
\bar{D}^{2}\bar{\Phi}^ID^{2}\Phi_I \right) \right)
,\nonumber\\
B1&=& 6G_0(p_i^2)\sum_I^3\mbox{Tr} \left( \bar{\Phi}^I\Phi_I \right),\\
B2&=&6\left(-G_0(s_{12})\sum_I^3\mbox{Tr} \left( \bar{\Phi}^I\Phi_I
\right) + G_1^{\alpha\dot{\beta}}(s_{12})\sum_I^3 \mbox{Tr} \left(
\bar{D}^{\dot{\beta}}\bar{\Phi}^ID^{\alpha}\Phi_I \right) \right.
\nonumber
\\ && \left. +
G_1(s_{12})\sum_I^3\mbox{Tr} \left(
\bar{D}^{2}\bar{\Phi}^ID^{2}\Phi_I \right) \right) ,\nonumber \eeqa
and again selecting the proper structures, prior to the application
of the on-shell conditions, gives \beq
\mathcal{M}^{(1)}=C1(s_{12},p_1^2,p_2^2)+B1(p_1^2)+B1(p_2^2)+2B2(s_{12})=6(s_{12}-p_1^2-p_2^2)
G_1(s_{12})-6G_0(s_{12}),\eeq

The UV divergent part of the answer is given by $6G_0$ and
extracting the coefficient of the $1/\epsilon$ pole, which is the
first coefficient in the anomalous dimension expansion
$\gamma_{\mathcal{K}}(\lambda)=\gamma_{\mathcal{K}}^{(1)}\lambda+
\ldots $ , we obtain the well-known result \beq
\gamma_{\mathcal{K}}^{(1)}=\frac{3}{8\pi^2}. \eeq

We see that up to one loop all the form factors for the
operators
$\mathcal{C}_{IJ},\mathcal{V}_{I}^{J},\mathcal{K},\mathcal{O}^{(n)}_I$
are proportional to $G_1$, the scalar triangle function (see App. B).

\subsection{$\mathcal{C}_{IJ}$ form factors at 2-loops}\label{sb32}

\begin{figure}[ht]\vspace{0.3cm}
 \begin{center}
 \leavevmode
  \epsfxsize=15cm
 \epsffile{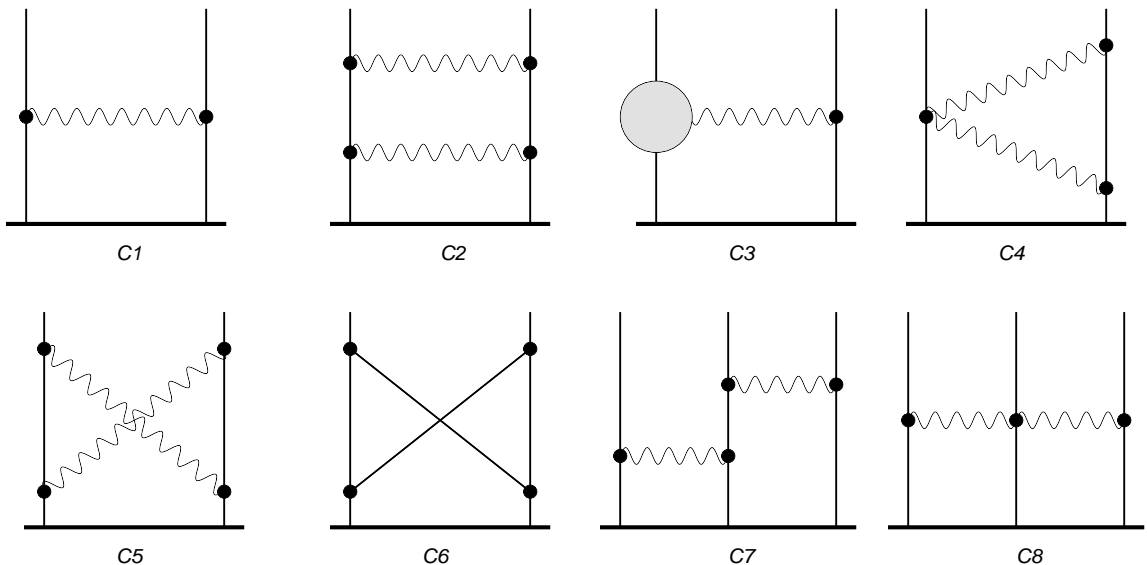}
 \end{center}\vspace{-0.2cm}
\caption{The relevant supergraphs in the chiral case. $C1$ is the
one-loop diagram, and the rest are two-loop ones. For the chiral
operator $\mathcal{C}_{IJ}$ with two legs the last two diagrams $C7$
and $C8$ do not exist,  they are only relevant for the operator
$\mathcal{O}_n$ with $n\geq 3$. A grey circle is the one-loop
effective vertex.}\label{supgr2}
 \end{figure}

We see that the form factors associated with $\mathcal{C}_{IJ}$ and
$\mathcal{V}_{I}^{J}$ are equal to each other at the one-loop level.
This is because $\mathcal{C}_{IJ}$ and $\mathcal{V}_{I}^{J}$ are
different components of the $\mathcal{N}=4$ conserved stress tensor.
In what follows we compute  the $\lambda^2$ contribution to
$\mathcal{M}$ for $\mathcal{C}_{IJ}$  since for the chiral operator
the  $D$-algebra is essentially simpler.  The corresponding diagrams
are shown in Fig.\ref{supgr2}.

Their contribution to the form factor are summarized in  Table \ref{chir}. All the relevant integrals are given in App.B
\begin{table}[htb]
\begin{center}
 \begin{tabular}{|c|c|c|}
   \hline
   $N$ & $\mathcal{C}_{IJ}$ & $\mathcal{O}_I^{(n)}$\\
   \hline
   $\textbf{C1}$ & $2s_{12}G_1(s_{12})$&$s_{ii+1}G_1(s_{ii+1})$\\
   \hline
   $\textbf{C2}$ & $4s^2_{12}G_2(s_{12})$&$s^2_{ii+1}G_2(s_{ii+1})$\\
   \hline
   $\textbf{C3}$ & $2s_{12}G_3(s_{12})+2s_{12}G_4(s_{12})$&$s_{ii+1}G_3(s_{ii+1})+s_{ii+1}G_4(s_{ii+1})$\\
   \hline
   $\textbf{C4}$ & $-6s_{12}G_3(s_{12})$&$-2s_{ii+1}G_3(s_{ii+1})$\\
   \hline
   $\textbf{C5}$ & $2G_5^a(s_{12})$ &$0$\\
   \hline
   $\textbf{C6}$ &$2G_5^b(s_{12})$ &$0$\\
   \hline
   $\textbf{C7}$ &$0$ &$(s_{ii+1}+s_{i+1i+2}+s_{ii+2})
   G_{6}(s_{ii+1},s_{i+1i+2},s_{ii+2})$\\
   \hline
   $\textbf{C8}$ &$0$ &$s_{i+1i+2}G_7(s_{ii+1},s_{i+1i+2},s_{ii+2})$\\   \hline
 \end{tabular}
 \caption{The contributions to the form factors from the individual diagrams}\label{chir}
 \end{center}
\end{table}

Adding all together and combining with the leading order one gets
\beq
\mathcal{M}^{(2)}=C2+C3+C4+C5+C6=2G_5^a+2G_5^b-4s_{12}G_3+2s_{12}G_4+4s^2_{12}G_2.
\eeq Using the identity
$$G_5^a+G_5^b=2s_{12}G_3-s_{12}G_4+\frac{s_{12}^2}{2}G_5$$
one can reduce it to $4s_{12}^2~G_2+s_{12}^2~G_5$ and finally get
\beq
\mathcal{M}=1+\lambda(2s_{12}~G_1)+\lambda^2(4s_{12}^2~G_2+s_{12}^2~G_5),
+ O(\lambda^3). \eeq

The general structure of form factors $\mathcal{M}$ "with two
external legs" in the gauge theory with zero beta-function has the
following form\footnote{
The IR exponentiation for two-leg form factors in QCD was established earlier in \cite{Magnea:1990zb}.}:

\beq\label{fact}
\log(\mathcal{M})=\frac{1}{2}\sum_{i=1}^{2}\left(\hat{M}(s_{i,i+1}/\mu^2)\right)+O(\epsilon).
\eeq Here we introduced \beq   \label{cup}
\hat{M}(s_{i,i+1}/\mu^2)=-\frac{1}{2}\sum_l\left(\frac{\lambda}{16\pi^2}\right)^l
\left(\frac{\Gamma^{(l)}_{cusp}}{(l\epsilon)^2}+\frac{G^{(l)}}{l\epsilon}+C^{(l)}\right)
\left(\frac{s_{i,i+1}}{\mu^2}\right)^{l\epsilon}, \eeq where
$\Gamma^{(l)}_{cusp}$ are the coefficients of perturbative expansion
of the cusp anomalous dimension $\Gamma_{cusp}(\lambda) =\sum_l
\Gamma^{(l)}_{cusp} \lambda^l$ which is a universal quantity that
governs the IR behavior of gauge theory amplitudes and the UV
behavior of the Wilson loops, and some local gauge invariant
operators.  $G^{(l)}$ are the coefficients of perturbative expansion
of the so-called collinear anomalous dimension
$G(\lambda)=\sum_lG^{(l)}\lambda^l$  and $C^{(l)}$ are some
constants. The quantities $G^{(l)}$ and $C^{(l)}$ are regularization
and scheme dependent. Performing the expansion of the integrals
$G_1,G_2,G_5$ in $\epsilon$ (see App. B) and introducing the
notation \beq  a \ = \ \frac{\lambda
}{16\pi^2}e^{-\epsilon\gamma_E}, \eeq where $\gamma_E$ is the
Euler--Mascheroni constant, one gets \beq
\log(\mathcal{M})=a\left(\frac{s_{12}}{\mu^2}\right)^{-\epsilon}\left(\frac{-2}{\epsilon^2}+\zeta_2\right)
+a^2\left(\frac{s_{12}}{\mu^2}\right)^{-2\epsilon}\left(\frac{\zeta_2}{\epsilon^2}+\frac{\zeta_3}{\epsilon}\right)+O(a^3)
\eeq where $\zeta_n$ are the Riemannian zeta functions $$\zeta_n =
\sum_{k=1}^\infty \frac{1}{k^n}.$$ From this answer and comparing
with eq.(\ref{cup}) we can extract the first two terms of
perturbative expansion over $a$ for the cusp and the collinear
anomalous dimensions and the finite terms \beqa
&&\label{cu} \Gamma^{(1)}_{cusp}=4,~\Gamma^{(2)}_{cusp}=-8\zeta_2, \\
&& G^{(1)}=0,~G^{(2)}=-\zeta_3, \\
&&C^{(1)}=-\zeta_2,~C^{(2)}=0.
\eeqa

Note that the maximal transcendentality principle holds which in our
case means that  if we attach to each logarithm and $\pi$ the level
of transcendentality equal to $1$ and to polylogarithms $Li_n(x)$
and $\zeta_n$ the level of transcendentality equal to $n$, then  at
the given order of perturbation theory the coefficient for the
$n$-th pole $1/\epsilon^n$ has the overall transcendentality  equal
to $2l-n$, where $l$ is the number of loops.  For a product of several factors it is given by the  sum of
transcendentalities  of each factor.

The leading IR behavior of $\mathcal{M}$ in this case can also be
captured by considering the Wilson line with one cusp
\cite{Korchemskaya:1992je} . So in this sense the dual description in
terms of Wilson loops for such form factors is well known.

One can see  that  the finite part for the form factor  is given
only in one loop and vanishes at two loops. However, this is a scheme
dependent result, and, for example, if we choose a different scheme
and replace $\exp(l\epsilon\gamma_E)$ for the $l$-th loop by
$\Gamma(1-\epsilon)^l$, we obtain in this scheme: \beq
\tilde{C}^{(1)}=0,~\tilde{C}^{(2)}=-\zeta_2^2,\eeq while the first
two coefficients in the perturbation theory for the cusp anomalous
dimension $\Gamma_{cusp}^{(1)}$ and $\Gamma_{cusp}^{(2)}$ remain the
same, which reflects the fact that they are scheme independent.

The same result  is true \cite{vanNeerven:1985ja} for  the finite part for the
form factor of a slightly different operator  $\mathcal{V}_X$  but belonging to the same
stress-tensor superconformal multiplet.

\section{Form factors with $\Delta_0=n$, $n>2$}\label{s4}
\begin{figure}[htb]
 \begin{center}
 \leavevmode
  \epsfxsize=3.6cm
 \epsffile{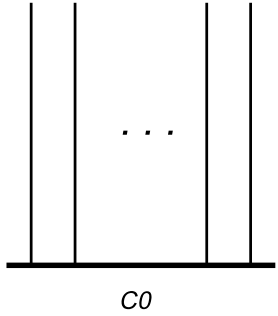}
 \end{center}\vspace{-0.2cm}
 \caption{The tree contribution to $\mathcal{O}^{(n)}_I$.}\label{1LoopNPoint}
\end{figure}

Here we present the results of calculation of  the form factors of
the chiral half-BPS operators $\mathcal{O}_I^{(n)}$ introduced
earlier. The tree-level contribution for the form factor is
presented on Fig. \ref{1LoopNPoint}.
In the first order of perturbation theory, similar to the
form factors of operators with conformal dimension $2$, the
contribution is given by the triangle type diagram and the
corresponding form factor, after performing the $D$-algebra and the
color algebra, is \beq \mathcal{M}^{(1)}=\sum_{i=1}^{n}
s_{i,i+1}~G_1, \eeq where we assume hereafter  $s_{n+i,n+i+1} \ = \
s_{i,i+1}.$

\subsection{$\mathcal{O}_I^{(n)}$, $n=3$ form factors at 2-loops}\label{sb41}

At the second order of perturbation theory the corresponding diagrams are shown in Fig.\ref{supgr2}
and their contributions are summarized in Table \ref{oper}.
\beqa
\mathcal{M}^{(2)}&=&\sum_{i=1}^{n}\left(s^2_{ii+1}G_2(s_{ii+1})-s_{ii+1}G_3(s_{ii+1})+s_{ii+1}G_4(s_{ii+1})\right)\nonumber \\
&+&\sum_{i=1}^{n}\left(s_{i+1i+2}G_7(s_{ii+1},s_{i+1i+2},s_{ii+2})+s_{ii+1}G_7(s_{i+1i+2},s_{ii+1},s_{ii+2})\right) \nonumber \\
&+&\sum_{i=1}^{n}\left(s_{ii+1}+s_{i+1i+2}+s_{ii+2}\right)G_{6}(s_{ii+1},s_{i+1i+2},s_{ii+2})
\eeqa

The next step is to establish the factorization property
(\ref{fact},\ref{cup}). Expanding the relevant  scalar integrals in
$\epsilon$ (see App. B), we obtain for $\log(\mathcal{M})$: \beq
\log(\mathcal{M}) =
\sum_{i=1}^3a\left(\frac{s_{ii+1}}{\mu^2}\right)^{-\epsilon}\left(-
\frac{1}{\epsilon^2} + \frac{\zeta_2}{2} \right) + \sum_{i=1}^3 a^2
\left(\frac{s_{ii+1}}{\mu^2}\right)^{-2\epsilon}\left(
\frac{\zeta_2}{2 \epsilon^2} + \frac{7 \zeta_3}{2 \epsilon} \right)
+ \mbox{fin.part.}\eeq

As in the case of the form factors of the operators with conformal dimension $2$, we can
extract the first two terms for the cusp and collinear anomalous
dimensions. This gives
 \beqa
&&\label{cusp} \Gamma^{(1)}_{cusp}=4,
\Gamma^{(2)}_{cusp}=-8\zeta_2, \\
&&\label{collin} G^{(1)}=0, G^{(2)}=-7
\zeta_3.\eeqa
Notice that the values of the cusp anomalous dimension $\Gamma^{(l)}$ are universal and coincide with
(\ref{cu}), while those of the collinear anomalous dimension depend on the form factor at hand.

We would like to emphasize the highly nontrivial cancelations between
the polylogarithms that occurred for $\log(\mathcal{M})$ for the whole set of scalar integrals
(the individual contributions to the poles from the scalar integrals are usually
 complicated polynomials of logarithms and polylogarithms of
different weight, see, for example, the expansions in $\epsilon$ of
$G_6$ and $G_{7}$ in App. B).

We see that the IR factorization property holds for the form
factors like for the amplitudes.

\subsection{$\mathcal{O}_I^{(n)}$ form factors for $n>3 $ at 2-loops}\label{sb42}

The corresponding contribution to the form factor up to
$\lambda^2$ is similar to the case of $n=3$ but has an additional
term coming from the factorized diagrams \beqa
\mathcal{M}^{(2)}&=&\sum_{i=1}^{n}\left(s^2_{ii+1}G_2(s_{ii+1})-s_{ii+1}G_3(s_{ii+1})+s_{ii+1}G_4(s_{ii+1})\right)\nonumber \\
&+&\sum_{i=1}^{n}\left(s_{i+1i+2}G_7(s_{ii+1},s_{i+1i+2},s_{ii+2})+s_{ii+1}G_7(s_{i+1i+2},s_{ii+1},s_{ii+2})\right) \nonumber \\
&+&\sum_{i=1}^{n}\left(s_{ii+1}+s_{i+1i+2}+s_{ii+2}\right)G_{6}(s_{ii+1},s_{i+1i+2},s_{ii+2})\nonumber \\
&+&\sum_{i=1}^{n}\sum_{j=1}^{n}s_{ii+1}G_1(s_{ii+1})s_{jj+1}G_1(s_{jj+1})
\eeqa

Performing the expansion over $\epsilon$  we obtain the logarithm of the
form factor up to the second order of perturbation theory
$\log(\mathcal{M})$ \beq \log(\mathcal{M}) =
\sum_{i=1}^na\left(\frac{s_{ii+1}}{\mu^2}\right)^{-\epsilon}\left(-
\frac{1}{\epsilon^2} + \frac{\zeta_2}{2} \right) + \sum_{i=1}^n a^2
\left(\frac{s_{ii+1}}{\mu^2}\right)^{-2\epsilon}\left(
\frac{\zeta_2}{2 \epsilon^2} + \frac{7 \zeta_3}{2 \epsilon} \right)
+ \mbox{Fin.part} .\eeq

The first two coefficients for the cusp and collinear anomalous
dimension which we can extract from the above expression coincide
with the coefficients obtained earlier for $n=3$,
(\ref{cusp}) and (\ref{collin}), respectively. As for the  finite part
\beq
\mbox{Fin.part.} = \lambda F^{(1)}(s_{12}, \ldots, s_{n1}) +
\lambda^2 F^{(2)}(s_{12}, \ldots, s_{n1}) + O(\lambda^3),\eeq
 at one loop it is  trivial $F^{(1)}=0$, and the two loop expression $F^{(2)}$,
contrary to the previous case,   is a complicated function
containing logarithms, polylogarithms and generalized Goncharov
polylogarithms \cite{Goncharov} of several variables.
All the relevant expressions can be found in App. B.

Note, the result is still much simpler than in the non-supersymmetric
case \cite{vanNeerven:1985xr} and the maximal transcendentality
principle still holds.

\subsection{Collinear Limit}

Here we restrict ourselves to the three-leg form factors for which we
can study the simplified kinematics and express the
finite part  in terms of logarithms only without
polylogarithms or Goncharov generalized polylogarithms.  The most
difficult part which appears in our calculation comes from the
diagram involving the interaction of three external fields. It reduces to the integral  $G_7$ which is
expressed in terms of the Appell function of two variables
$$F_1(1;2 \epsilon, 1; 2 + \epsilon |x,y)$$ and  after the
$\epsilon$-expansion one obtains the generalized Goncharov
polylogarithms \cite{Goncharov}. One can see from integral
representation of the Appell function \beq \label{AppLim1}
F_1(a;b_1,b_2;c | x, y) = \frac{\Gamma(c)}{\Gamma(a) \Gamma(c-a)}
\int_0^1 \frac{u^{a-1} (1-u)^{c-a-1}}{(1-u x)^{b_1} (1-u y)^{b_2}}
du, \mbox{Re} a, \mbox{Re} (c-a) > 0 \eeq that the only way to
achieve  the desired simplification is to have one of the arguments
equal to $0$ or to $1$. In the first case one gets \beq
\label{AppLim2} F_1(a;b_1,b_2;c | x, 0) \ = \ \, _2F_1(a, b_1; c |
x),\eeq and similarly in the second case  \beq F_1(a;b_1,b_2;c | x,
1) \ = \ \frac{\Gamma(c) \Gamma(c-a-b_2)}{\Gamma(c-a)\Gamma(c-b_2)}
\, _2F_1(a, b_1; c-b_2 | x).\eeq

Such a simplification can occur in two-dimensional kinematics when
one of the kinematical variables $s_{12}, s_{13}$ or $s_{23}$
equals  $0$.  The  other motivation for this kinematics is
the recent strong coupling calculations which have been performed for the
$AdS_3$ sub-manifold of $AdS_5$  which corresponds to the
degenerate $1+1$ kinematics in a dual theory \cite{Maldacena:2010kp}.

The $1+1$ dimensional kinematics necessarily contains a collinear
configuration of the space components $\vec{p}_i$ of momenta $p_i$.
For massless gauge theory it is known that in such collinear limit
the factorization of the IR divergencies fails. For the partial
color ordered amplitudes in collinear limit  when two momentums
$p_i$ and $p_{i+1}$ are replaced by $zp$ and $(1-z)p$ the deviation
from the factorized form  is governed by the so-called ''loop
splitting functions" $r^{(l)}_s(\epsilon,z,p^2)$, $l$ being the
number of loops. In the $\mathcal{N}=4$ SYM theory
$r^{(l)}_s(\epsilon,z,p^2)$ have an  iterative structure, so one can
write the following relation valid  in collinear limit (see, for
example, the discussion in \cite{bds05})
$$
\log(M_n) \to
\frac{1}{2}\hat{M}_{n-1}+\sum_l\lambda^l\Gamma^{(l)}_{cusp}r^{(l)}_s(l\epsilon,z,p^2)
+\sum_l\lambda^lF^{(l),~coll}_{n-1}+O(\epsilon)
$$
$$
r^{(l)}_s(l\epsilon,z,p^2) \sim
\frac{1}{\epsilon^2}\left(\frac{p^2}{\mu^2}\right)^{\epsilon}
\left(-\frac{\pi\epsilon}{\sin(\pi\epsilon)}\left(\frac{1-z}{z}\right)^{\epsilon}
+2\sum_{k=0}\epsilon^{2k+1}Li_{2k+1}(\frac{-z}{1-z})\right)
$$

We expect that similar violation of the IR factorization happens
in the case of the form factors. Indeed, in the $s_{23} \to 0$ limit we
have, up to $\lambda^2$ \beqa && \log(\mathcal{M}) =  \sum_{i=1}^{2}
a\left(\frac{s_{ii+1}}{\mu^2}\right)^{-\epsilon}\left(-
\frac{1}{\epsilon^2} + \frac{\zeta_2}{2} \right)+ \sum_{i=1}^2
a^2\left(\frac{s_{ii+1}}{\mu^2}\right)^{-2\epsilon} \left(
\frac{\zeta_2}{2\epsilon^2} + \frac{\zeta_3}{2 \epsilon} \right)
\\ \nonumber &&  \makebox[-2em]{}
+\sum_{i=1}^{2}a^2\left(\!\frac{s_{ii+1}}{\mu^2}\!\right)^{-2\epsilon}\!\!\left(\!\frac{-6\zeta_2+
3 \log^2
\frac{s_{12}}{s_{13}}}{96\epsilon^2}+\frac{19\zeta_3}{8\epsilon}\!\right)\!
-\!\frac{a^2}{2880} \left(\! 75 \log^4 \frac{s_{12}}{s_{13}} \!+\! 120\pi^2
\log^2 \frac{s_{12}}{s_{13}} \!-\!317 \pi^4\! \right).\eeqa

\section{Dual conformal invariance}\label{s5}

Here we would like to discuss the property of dual conformal
invariance of the integrals appearing in our calculation. It is
believed that all the integrals entering into the calculation of the
amplitudes should be dual conformal invariant when external legs are
off-shell \cite{dualKorch}. This means that the answers should be
expressed in terms of the conformally invariant cross-ratios,  which
restricts the form of the result.  On mass-shell this dual conformal
symmetry has an anomaly but still remains a very important ingredient
in understanding  the properties of the on-shell amplitudes and the
Wilson loops (see, for example
\cite{BeisertYangianAmpl,Alday:2008yw}).

When calculating the form factors one has similar integrals though
they do not possess explicit dual conformal invariance. However, it
is remarkable that all the integrals that appear in our calculation
can be obtained from those contributing to the amplitudes by some
reduction which we describe below. Because of this  connection they
also bear some conformal properties.

Consider several examples.  At one loop there is a single triangle
diagram contributing to all the form factors. The one-loop triangle
is the first in a chain of  the ladder type diagrams
\cite{Usyukina:1992jd} and  has the property of dual conformal
invariance \cite{Usyukina:1992jd,Drummond:2006rz}. This diagram is
to be connected to the box diagram, which is dual
conformal\cite{Drummond:2006rz}, in the following  way. Consider the
one-loop off-shell box diagram in momentum space which is given by
the integral \beq \label{box}D^{1-loop}(p_1,p_2,p_3,p_4) =
\int\frac{d^Dk}{(2\pi)^D}\frac{1}{k^2(k-p_1)^2(k+p_2)^2(k+p_2+p_3)^2}.
\eeq Introducing the dual coordinates $x_i$ as
$$
p_1=x_{12},~p_2=x_{23},~p_3=x_{34},~p_{4}=x_{41},~k=x_{5},
$$
we can rewrite the initial integral in the following form  \beq
\label{box} D^{1-loop}(x_1,x_2,x_3,x_4) = \int \frac{d^D
x_5}{x_{15}^2 x_{25}^2 x_{35}^2 x_{45}^2} =
\frac{1}{x_{13}^2x_{24}^2} \Phi(X,Y),\eeq where we introduced the
notation $x_{ij} = x_i - x_j$ and $\Phi(X,Y)$ is the function given
in \cite{Usyukina:1992jd}, $X$ and $Y$ are the conformal
cross-ratios
$$X \ = \ \frac{x_{12}^2x_{34}^2}{x_{13}^2x_{24}^2}, \ Y \ = \
\frac{x_{14}^2x_{23}^2}{x_{13}^2x_{24}^2}.$$

If we now multiply (\ref{box}) by $x_{12}^2$ and take the limit $x_2
\rightarrow \infty$,  we obtain  the one-loop triangle diagram
\cite{Drummond:2006rz}
 \beq C^{1-loop} \ = \ \lim_{x_2 \rightarrow \infty} x_{12}^2
\int \frac{d^D x_5}{x_{15}^2 x_{25}^2 x_{35}^2 x_{45}^2} = \int
\frac{d^4 x_5}{x_{15}^2 x_{35}^2 x_{45}^2} \ = \ \frac{1}{x_{34}^2}
\Phi(x,y),  \label{trig}
\eeq with $$x \ = \ \frac{x_{34}^2}{x_{13}^2}, \ y \ = \
\frac{x_{14}^2}{x_{13}^2}.$$

\begin{figure}[htb]
 \begin{center}
 \leavevmode
  \epsfxsize=9cm
 \epsffile{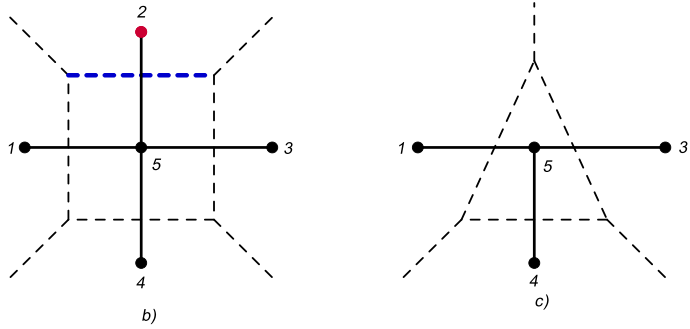}
 \end{center}\vspace{-0.2cm}
 \caption{The one-loop triangle diagram from the one-loop box
diagram. The red dot should be taken to infinity, and the blue line
(propagator in momentum space) should be contracted to a
point.}\label{triangle1}
\end{figure}

Schematically, the described procedure of obtaining  the one-loop
triangle diagram from the one-loop box diagram is represented in
Fig. \ref{triangle1}.
On the left hand side  one has the one-loop
box diagram together with the dual grid, the black lines represent
the denominator terms appearing in the integral in $x$--space.
Taking the limit $x_2 \rightarrow \infty$ in (\ref{trig}) is
equivalent to removing the grid line $x_{25}$ from the dual graph
and shrinking the crossed line to a point in the initial graph. The
resulting  initial  graph corresponds to the triangle diagram,  as is
shown on the right hand side.  This way  the triangle diagram can be
obtained from the box one and inherit its property of dual conformal
invariance.

\begin{figure}[htb]\vspace{0.3cm}
 \begin{center}
 \leavevmode
  \epsfxsize=9cm
 \epsffile{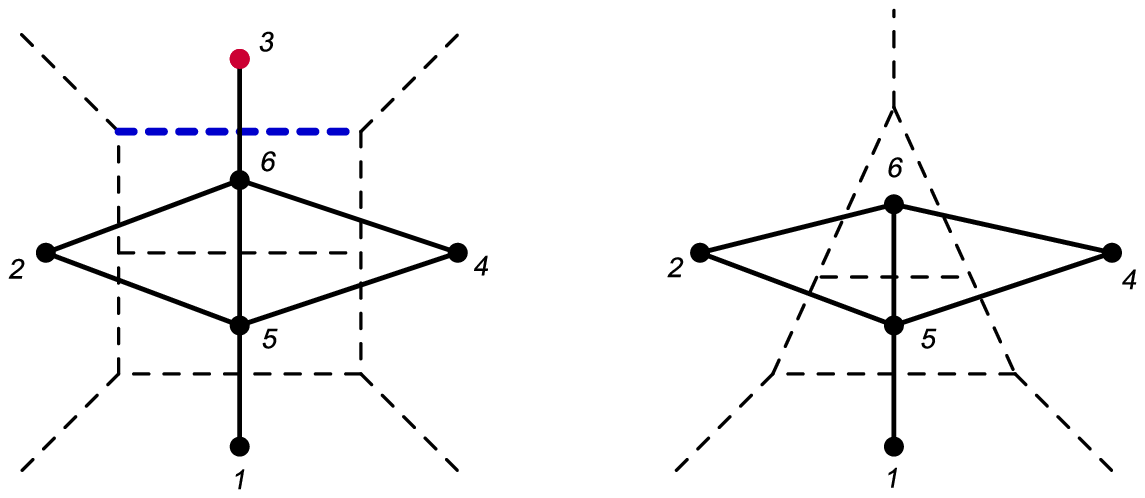}\vspace{0.3cm}

 \epsfxsize=9cm
 \epsffile{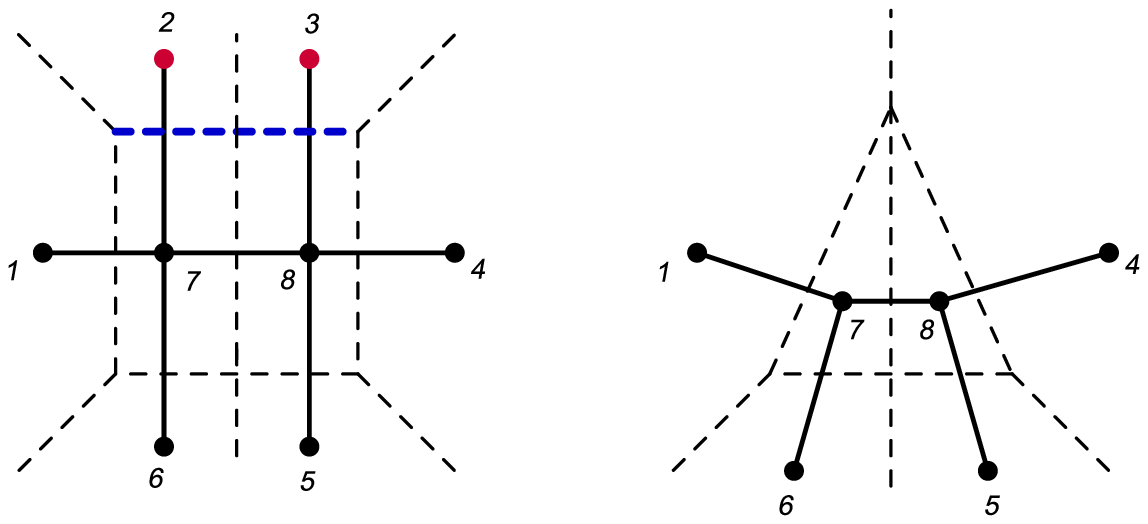}\vspace{0.2cm}

  \epsfxsize=9cm
 \epsffile{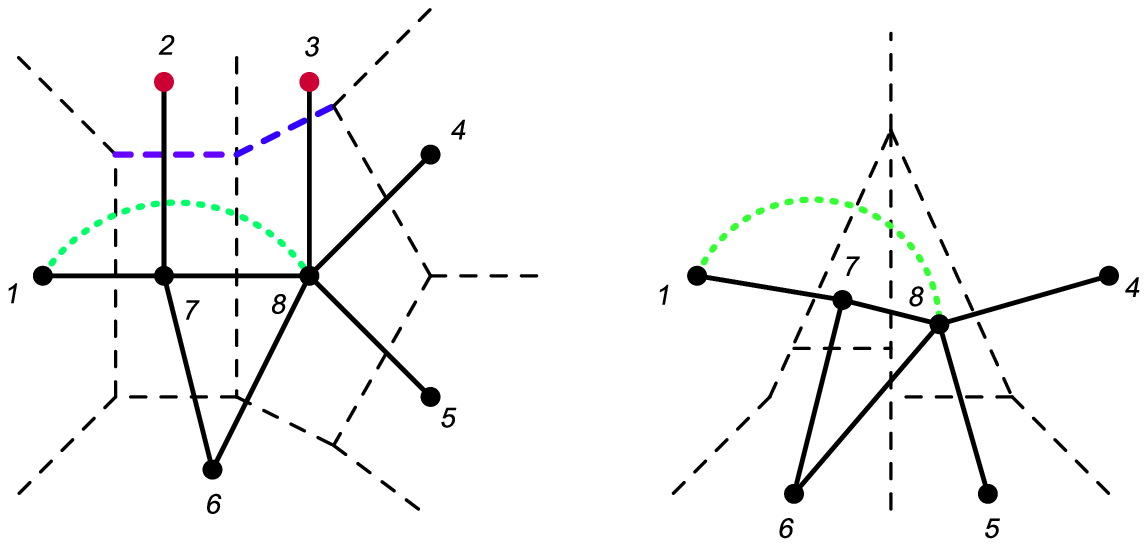}
  \caption{The two-loop ladder type triangle diagrams from the two-loop
box diagrams. Green arc corresponds to the
presence of a numerator}\label{VBox}
\end{center} \end{figure}

In the same manner one can
show how the other diagrams  which appear in our calculation can be obtained from the corresponding
diagrams  entering into the amplitude calculations.  Schematically, we present this procedure  in Fig.\ref{VBox}.

For the vertical box diagram   one has to take the limit $x_3
\rightarrow \infty$. As in the previous case this corresponds to
removing the grid line $x_{36}$ (and shrinking the corresponding
crossed line) which results in the diagram shown on the right hand
side. This is exactly the ladder integral that appears in two-loop
calculation of the form factor with $n \geq 2$ legs.

For the  horizontal box diagram one should take the combined limit
$x_2, x_3 \rightarrow \infty$ which is schematically shown on the
right hand side.  This is the new type of integrals which appears
only in the case when  $n>2$.

The same procedure is expected to work at higher levels of
perturbation theory. Our conjecture is that the integrals appearing
at any order of perturbation theory are obtained from  dual
conformal invariant diagrams by contraction of $n$ propagators at
the $n$-th loop order.

\section{Discussion}\label{s6}

In this paper we continue the perturbative study of the form factors
at weak coupling for  the $\mathcal{N}=4$ SYM theory which was initiated
in \cite{vanNeerven:1985ja} where the author considered the form
factor for the operator $\mathcal{V}_X$ of conformal dimension $2$
which in $\mathcal{N}=1$ superspace notations takes the form $2
\mbox{Tr} \Phi_1 \overline{\Phi}_1 - \mbox{Tr} \Phi_2
\overline{\Phi}_2 - \mbox{Tr} \Phi_3 \overline{\Phi}_3$.  The
original calculation has been performed in components and the form
factor was computed up to the second order of perturbation theory.
In our paper, we started with the operators belonging to the
stress-tensor superconformal multiplet, namely, with
$\mathcal{V}_{I}^{J}=\mbox{Tr}\left(\bar{\phi}^{J}\phi_I\right)$ and
$\mathcal{C}_{IJ}=\mbox{Tr}\left(\phi_I\phi_J\right)$ and calculated
them up to the first and second order of perturbation theory,
respectively. We obtained the same results as in
\cite{vanNeerven:1985ja} as it was expected.

Then we considered  the Konishi operator $\mathcal{K}=\sum_{I}
\mbox{Tr} \left(\bar{\phi}^I \phi_I \right)$ with classical
conformal dimension $2$  in the one loop approximation. Not being
protected by supersymmetry this operator has the UV divergences
which have to be renormalized.

The main result of our paper is the calculation of the two-loop
 form factors for the half-BPS operators
$\mathcal{O}_I^{(n)}=\mbox{Tr}\left(\phi_I^n\right), n>2$. At the one
loop level the answer for the form factor is very simple given by
 triangle diagram while at  two-loops it is essentially more
complicated. The analytical expressions for the two-loop results
are given in terms of the Gauss hypergeometric
functions and the Appell function of two variables. Their expansion over
$\epsilon$  up
to $O(\epsilon)$ leads to logarithms, polylogarithms and,  because of the Appell function,
generalized Goncharov polylogarithms of several variables.  However, all of them have
 the same transcendentality
\cite{Transc,KotTransc}.

In the simplified kinematics the answers become much more
simple. Thus, in two-dimensional (or
$1+1$-dimensional) kinematics for the form factors of the half-BPS
operators $\mathcal{O}_I^{(3)}$  it is possible to get rid
of the Appell functions and after expanding over $\epsilon$ to get the result in terms
of the ordinary logarithms.

For all the considered form factors we observe the factorization of
the IR divergences up to the second order of perturbation theory.
This allows us to derive the first two terms of expansion for the cusp
anomalous dimension in coincidence  with the other
calculations and for the collinear anomalous dimension, where we obtained
the first
nontrivial coefficient at two loops $G^{(2)} \ = \ -7 \zeta_3$.  It
differs from the collinear anomalous dimension coming from the
amplitude calculation but coincides with
collinear anomalous dimension for the light-like Wilson loop
\cite{Alday:2008yw,Korchemskaya:1992je}.

The remarkable  part of our calculation besides factorization is the
fact that the one- and two-loop integrals contributing to  the form
factors of the operators $\mathcal{O}_I^{(n)}, n>2$ are related to
the dual conformal invariant integrals appearing in the calculation
of the amplitudes. One has to look at the ''parent" integral which
appears while considering the amplitudes and shrink $n$ propagators
at the $n$-th order of perturbation theory. This dual conformal
invariance together with the original conformal invariance might
lead to a wider algebra eventually constraining the form of the
answer and reveal the integrability property of a theory. It is
important whether the powerful $\mathcal{N}=4$ covariant on-shell
methods such as recurrent relations (see recent \cite{He:2010ju} for
example) can be generalized for the form-factors studied in our
paper.

Note added: while finishing writing the paper we became aware of the paper which is closely connected
to the subject studied here \cite{Brandhuber:2010ad}.

\section*{Acknowledgements}

We would like to thank N. Beisert, A. Gorsky, A. Grozin, L. Lipatov,
T. McLoughlin, V. Smirnov and A. Zhiboedov for valuable discussions.
We thank T. Huber for pointing out several typos in the first
version of our paper. GV would like to thank TPI (Minnesota) for
hospitality during his visit in November 2010 while finishing the
paper. Financial support from RFBR grant \# 08-02-00856 and the
Ministry of Education and Science of the Russian Federation grant \#
1027.2008.2 is kindly acknowledged.

\appendix
\section{Feynman rules in $\mathcal{N}=1$ superspace}

We want to present here  the  essential elements of
$\mathcal{N}=1$ superspace technique relevant to our computations.

In terms of $\mathcal{N}=1$ superfields the $\mathcal{N}=4$ SYM
action can be rewritten as (hereafter we use the notation of
\cite{1001}, see recent examples of application of the same
technique in \cite{SiegSuperspacenew,Penati:1999ba,N=4Realesys})
\beq S^{\mathcal{N}=4}= \int d^8z \mbox{Tr}\left(e^{-gV}\bar \Phi^I
e^{gV} \Phi_I\right) +\frac{1}{2g^2}\int d^6z \mbox{Tr} (W^\alpha
W_\alpha) +ig\int d^6z
\mbox{Tr}\left(\Phi_1[\Phi_2,\Phi_3]\right)+c.c., \label{lag}\eeq
where the superfield strength tensor $W_\alpha= \bar
D^2(e^{-gV}D_\alpha e^{gV}),$ $V=V^{a}T_a$ is the real
$\mathcal{N}=1$ vector superfield and $\Phi_I=\Phi_I^aT_a$ with
$I=1,2,3$ are the three chiral superfields ($I$ is the index of the
$SU(3)$ subgroup of $SU(4)_R$), $T_a$ are the generators of the
gauge group $SU(N_c)$ in adjoint representation. For performing
$SU(N_c)$ $T$-matrix manipulations we used FeynCalc package for
Mathematica \cite{FeynCalc}. The following normalization for $T^a$
is used in which the quadratic Casimir operator \beq
\mbox{Tr}(T^aT^b)=k_2\delta^{ab},~k_2=1/2. \eeq

The relevant Feynman rules for the propagators and vertices are
\beqa  \langle
V^aV^b\rangle&=&-\frac{1}{k_2}\delta^{ab}\frac{\delta_{12}}{p^2},
\nonumber \\  ~\langle
\bar{\Phi}_I^a\Phi_J^b\rangle&=&\frac{1}{k_2}\delta^{ab}\delta_{IJ}\frac{\delta_{12}}{p^2},\nonumber\\
V(\bar{\Phi}V\Phi)&=&igk_2f_{abc}\delta^{IJ}\bar{\Phi}^{a}_IV^{b}\Phi^{c}_J, \\
V(\Phi\Phi\Phi)&=&\frac{-g}{3!}\epsilon^{IJK}f_{abc}\Phi^a_I\Phi^b_J\Phi^c_K,\nonumber
\\
~V(\bar{\Phi}\bar{\Phi}\bar{\Phi})&=&\frac{-g}{3!}\epsilon^{IJK}f_{abc}\bar{\Phi}^a_I\bar{\Phi}^b_J\bar{\Phi}^c_K, \nonumber\\
V(\bar{\Phi}VV\Phi)&=&\frac{g^2}{2}k_2\delta^{IJ}f_{adm}f_{bcm}V^a\Phi^d_IV^b\bar{\Phi}^c_J. \nonumber
\eeqa
 where
$\delta_{12}=\delta^4(\theta_1-\theta_2)$ is the Grassmannian delta
function.

The effective one-loop triple vertex is given by \beq
V(\bar{\Phi}V\Phi)_{1-loop}=ig\frac{\lambda}{4}k_2f_{abc}\bar{\Phi}^a_I(-q)\Phi^b_J(-p)\hat{\mathcal{D}}V^c(p+q)
\int\frac{d^Dk}{(2\pi)^D}\frac{1}{k^2(k-q)^2(k+p)^2}, \eeq where
\beq
\hat{\mathcal{D}}=4D^{\alpha}\bar{D}^2D_{\alpha}+(p-q)^{\alpha\dot{\alpha}}[D_{\alpha},\bar{D}_{\dot{\alpha}}].
\eeq
As usual, the  vertex with $n$ chiral (anti-chiral) lines requires
additional $n-1$ $~\bar{D}^2$ (for anti-chiral $D^2$) acting on
chiral (anti-chiral) lines (or $n-1-m$ $~\bar{D}^2$ (for anti-chiral
$D^2$) if $m$ lines are external). We used SusyMath package for
Mathematica \cite{Susymath} for performing $D$-algebra for
supergraphs.

Traces in this case are taken over $\sigma$ matrices and are
evaluated in $D=4$ because dimensional reduction is used. The
following set of identities is useful: \beqa &&
\sigma^m=(\sigma^m)_{\alpha\dot{\beta}} ~
\bar{\sigma}^m=(\bar{\sigma}^m)^{\alpha\dot{\beta}} \nonumber \\ &&
p_{\alpha\dot{\beta}}=p_{m}(\sigma^m)_{\alpha\dot{\beta}}~
\bar{p}^{\alpha\dot{\beta}}=p_{m}(\bar{\sigma}^m)^{\alpha\dot{\beta}}
\nonumber \\ &&
\mathbf{1}=\delta^{\alpha}_{\beta},~\bar{\mathbf{1}}=\delta^{\dot{\alpha}}_{\dot{\beta}}
\nonumber \\ &&
\mbox{Tr}[\textbf{1}]=\mbox{Tr}[\bar{\mathbf{1}}]=\frac{D}{2}, \eeqa
where $D/2=2$ in dimensional reduction and also we have \beqa &&
\sigma^{m}\bar{\sigma}^{n}+\sigma^{n}\bar{\sigma}^{m}=-\eta^{mn}\textbf{1},
\nonumber \\ &&
\bar{\sigma^{m}}\sigma^{n}+\bar{\sigma^{n}}\sigma^{m}=-\eta^{mn}\bar{\textbf{1}}.
\eeqa

\section{Scalar integrals and their $\epsilon$ expansion}

\begin{figure}[h]
 \begin{center}
 \leavevmode
  \epsfxsize=13cm
 \epsffile{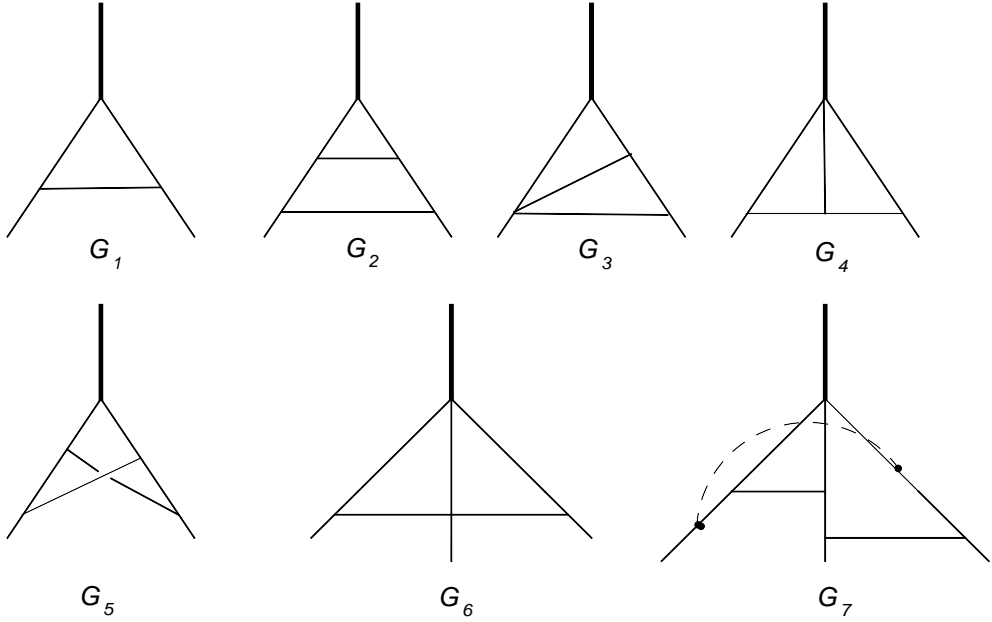}
 \end{center}\vspace{-0.2cm}
 \caption{The set of scalar integrals. The arc line in $G_7$ corresponds to the presence
 of the numerator $(k-p)^2$. Thick black line corresponds to off-shell leg with momentum $q$.
 All the other legs are on-shell.}\label{4}
 \end{figure}

Here we present the list of scalar integrals which we encountered in
our computation  shown in Fig. \ref{4}. All the integrals are
evaluated in $D=4-2\epsilon$ dimensions. For each loop the factor
$e^{\epsilon\gamma_E}$ is added in the integration measure, we also
do not write $4\pi$ which always appear with $\mu^2$.
\beqa
G_0&=&\int\frac{d^Dk}{(2\pi)^D}\frac{1}{k^2(k+p)^2}=
\left(\frac{e^{-\epsilon\gamma_E}}{16\pi^2}\left(\frac{p^2}{\mu^2}\right)^{-\epsilon}\right)
\left(\frac{1}{\epsilon}+2+O(\epsilon)\right), \eeqa
 \beqa
G_1&=&\int\frac{d^Dk}{(2\pi)^D}\frac{1}{k^2(k+p_1)^2(k+p_2)^2}
 \\ && \makebox[2em]{} =
-\left(\frac{e^{-\epsilon\gamma_E}}{16\pi^2}\left(\frac{s_{12}}{\mu^2}\right)^{-\epsilon}\right)
\frac{1}{s_{12}}\left(\frac{1}{\epsilon^2}-\frac{\zeta_2}{2}-\frac{7\zeta_3}{3}\epsilon
-\frac{47\pi^4}{1440}\epsilon^2+O(\epsilon^3)\right),\nonumber\\
G_2&=&\int\frac{d^Dk}{(2\pi)^D}\frac{d^Dl}{(2\pi)^D}\frac{1}{k^2l^2(k-p_1)^2(k+p_2)^2(k+l-p_2)^2(k+l-p_1)^2},
 \\ && \makebox[2em]{} =
\left(\frac{e^{-\epsilon\gamma_E}}{16\pi^2}\left(\frac{s_{12}}{\mu^2}\right)^{-\epsilon}\right)^2\frac{1}{s_{12}^2}
\left(-\frac{1}{4\epsilon^4}-\frac{5\pi^2}{24\epsilon^2}-\frac{29\zeta_3}{6\epsilon}-\frac{3\pi^4}{32}
+O(\epsilon) \right),\nonumber\\
G_3&=&\int\frac{d^Dk}{(2\pi)^D}\frac{d^Dl}{(2\pi)^D}\frac{1}{k^2l^2(k-l)^2(k+l-p_1)^2(k+l+p_2)^2(p_2+l)^2},
\\ && \makebox[2em]{} =
\left(\frac{e^{-\epsilon\gamma_E}}{16\pi^2}\left(\frac{s_{12}}{\mu^2}\right)^{-\epsilon}\right)^2\frac{1}{s_{12}}
\left(-\frac {1} {2 \epsilon^4}  + \frac {29 \zeta_3} {6 \epsilon} +
\frac {49 \pi ^4} {720}+O(\epsilon)\right), \nonumber\\
G_4&=&\int\frac{d^Dk}{(2\pi)^D}\frac{d^Dl}{(2\pi)^D}\frac{1}{k^2l^2(l+p_1)^2(k+p_2)^2(k-l+p_2)^2(l-p_2)^2}
\\ && \makebox[2em]{} =
\left(\frac{e^{-\epsilon\gamma_E}}{16\pi^2}\left(\frac{s_{12}}{\mu^2}\right)^{-\epsilon}\right)^2\frac{1}{s_{12}}
\left(\frac {1} {4 \epsilon^4} - \frac {\pi ^2} {24 \epsilon^2} -
\frac {8 \zeta_3} {3 \epsilon}- \frac {19 \pi ^4} {480} +O(\epsilon)
\right),\nonumber \\
G_5&=&\int\frac{d^Dk}{(2\pi)^D}\frac{d^Dl}{(2\pi)^D}\frac{1}{k^2l^2(k-l)^2(k-p_2)^2(k-l-p_1)^2(l-p_2-p_1)^2}
 \\ && \makebox[2em]{} =
\left(\frac{e^{-\epsilon\gamma_E}}{16\pi^2}\left(\frac{s_{12}}{\mu^2}\right)^{-\epsilon}\right)^2\frac{1}{s_{12}^2}
\left(-\frac {1} {\epsilon^4} + \frac {\pi ^2} {\epsilon^2}  + \frac
{83 \zeta_3} {3 \epsilon} + \frac {59 \pi ^4} {120}
+O(\epsilon)\right)\nonumber.\eeqa

The other integrals entering into the calculations are
\beqa
G_5^a&=&\int\frac{d^Dk}{(2\pi)^D}\frac{d^Dl}{(2\pi)^D}\frac{s_{12}k^2-\mbox{Tr}(\bar{p}_1p_2\bar{l}k)+\mbox{Tr}(\bar{p}_1k\bar{l}p_2)}{k^2l^2(k-l)^2(k-p_2)^2(k-l-p_1)^2(l-p_2-p_1)^2},\\
G_5^b&=&\int\frac{d^Dk}{(2\pi)^D}\frac{d^Dl}{(2\pi)^D}\frac{s_{12}k^2-\mbox{Tr}(\bar{p}_1p_2\bar{k}l)}{k^2l^2(k-l)^2(k-p_2)^2(k-l-p_1)^2(l-p_2-p_1)^2},\\
G_1^{\alpha\dot{\beta}}&=&\int\frac{d^Dk}{(2\pi)^D}\frac{k^{\alpha\dot{\beta}}}{k^2(k+p_1)^2(k+p_2)^2}.
\eeqa
$G_0$ is the scalar bubble integral, $G_1$ is the scalar triangle integral.
$G_2$, $~G_3$ $~G_4$ can be computed by means of the MB representation
as series in $\epsilon$ or to all orders in $\epsilon$ by means of the
differential equation technique\footnote{This integrals can be reduced to the set of
master topologies presented in \cite{Gehrmann:1999as}.}. $G_5$ can be
computed by means of the MB representation as series in $\epsilon$,
 the answer to all orders in $\epsilon$ is given in \cite{GehrmannCross}.
$G_6$ and $G_{7}$ can be computed by direct
evaluation of integrals over the Feynman parameters in terms of the
hypergeometric function $_2F_1$ and the Appell function $F_1$. The formulas from
\cite{Anastasiou:1999ui} and
\cite{Ellis:2007qk} were useful in verification of our computation.
$G_{6}$  can also be evaluated by means of the differential equation
technique \cite{Gehrmann:1999as}, the result coincides with ours
after the rearrangement of hypergeometric functions. Using the
notation $s_{12}=s,s_{14}=t,s_{13}=u$ the answers for $G_6$
 can be written as:
$$
c_{\Gamma}=\frac{\Gamma^3(1-\epsilon)\Gamma(1+2\epsilon)}{\Gamma(1-3\epsilon)}
$$
\begin{eqnarray}
G_{6}&=& \int \frac{d^Dk}{(2\pi)^D}\frac{d^Dl}{(2\pi)^D} \frac{1}{l^2(l-k)^2(l-p_1)^2(k+p_2)^2\ (k+p_2+p_3)^2}\nonumber\\
&=&\frac{e^{-2\epsilon\gamma_E}c_{\Gamma}}{(16\pi^2)^2}~\frac{1}{2\epsilon^3}\frac{1}{(1-2\epsilon)}\nonumber\\
&&\left\{\left(\frac{\mu^2}{t}\right)^{2\epsilon}\frac{1}{s}~_2F_{1}(1,1-2\epsilon,2-2\epsilon,-\frac
us)
 +\left(\frac{\mu^2}{s}\right)^{2\epsilon}\frac{1}{t}~_2F_{1}(1,1-2\epsilon,2-2\epsilon,-\frac ut)\right.\nonumber\\
&&\left. -\left(\frac{\mu^2}{s+t+u}\right)^{2\epsilon}\frac{s+t+u}{t
s}~_2F_{1}(1,1-2\epsilon,2-2\epsilon,-\frac{u(s+t+u)}{st})\right\}.
\end{eqnarray}
where the hypergeometric function is given by the following
expansion:(we used the Nested Sums computational
tool \cite{Moch} and HypExp package for Mathematica \cite{Huber:2005yg})
 \beqa && \, _2F_1 \left(
1-2\epsilon,1-2\epsilon;2-2\epsilon;x\right)
=\sum_{n=0}^3\epsilon^n~a_{n}(x)+O(\epsilon^4)
 \eeqa
$$
a_0(x)=-\frac{\log (1-x)}{x}
$$
$$
a_1(x)=+\frac{\left(2 Li_2(x)-(\log (1-x)-2) \log (1-x)\right)}{x}
$$
\begin{eqnarray*}
a_2(x)&=& -\frac{2}{3x} \left(\log ^3(1-x)-3 \log (x) \log
^2(1-x)-3 \log ^2(1-x)+\pi ^2 \log (1-x) \right.\\
&& \left. \makebox[3em]{} -6 (\log (1-x)-1) Li_2(x)-6 Li_3(1-x)-6
Li_3(x)+6 \zeta (3)\right)
\end{eqnarray*}
\begin{eqnarray*}
a_3(x)&=&-\frac{2}{3 x} \left(\log ^4(1\!-\!x)-6 \log (x) \log
^3(1\!-\!x)-2 \log ^3(1-x)+6 \log (x) \log ^2(1\!-\!x) \right. \nonumber \\
\nonumber && \left. \makebox[0em]{} +2 \pi ^2 \log ^2(1\!-\!x)-6 (\log
(1\!-\!x)-2) Li_2(x) \log (1\!-\!x)-12 Li_3(x) \log (1\!-\!x) \right. \nonumber \\
\nonumber && \left. \makebox[0em]{} +12 \zeta (3) \log (1-x)-2 \pi
^2 \log (1\!-\!x)\!-\!12 (\log (1\!-\!x)\!-\!1) Li_3(1\!-\!x)+12 Li_3(x) \right. \nonumber \\
\nonumber && \left. \makebox[0em]{} +12
Li_4\left(\frac{x}{x-1}\right)-12 \zeta (3)\right)
\end{eqnarray*}
The finite part of $G_{6}$ is then given by the following
expression:
\begin{eqnarray*}
(G_{6})_{fin}=\frac{1}{(16\pi^2)^2}~\frac{1}{2}\left\{\frac{t}{s}~a_3(-\frac
us)+~a_3(-\frac ut)
-\frac{s+t+u}{s}~a_3(-\frac{u(s+t+u)}{st})\right\}.
\end{eqnarray*}

In the case of $"1+1"$ dimensional kinematics the following limiting expressions for
$G_{6}$ are used:
\begin{eqnarray*}
G_{6}|_{t=0}&=&0, \\
G_{6}|_{s=0}&=&\frac{e^{-2\epsilon\gamma_E}c_{\Gamma}}{(16\pi^2)^2}~\frac{1}{4\epsilon^4}\left\{\frac{1}{(t+u)^{2\epsilon}}-\frac{1}{t^{2\epsilon}}\right\}\\
G_{6}|_{u=0}
&=&\frac{e^{-2\epsilon\gamma_E}c_{\Gamma}}{(16\pi^2)^2}~\frac{-1}{2\epsilon^3}\frac{1}{1-2\epsilon}
\left\{\frac{t+s}{s}\frac{1}{(t+s)^{2\epsilon}}- \frac ts
\frac{1}{t^{2\epsilon}}-\frac{1}{s^{2\epsilon}}\right\}
\end{eqnarray*}

For $G_7$ one has
\begin{eqnarray}
G_7&=&\int \frac{d^Dk}{(2\pi)^D}\frac{d^Dl}{(2\pi)^D} \frac{(k-p_1)^2}{k^2l^2(l-k)^2(l-p_1)^2(k+p_2)^2\ (k+p_2+p_3)^2}=G_{6}+
\nonumber\\ &+&\frac{e^{-2\epsilon\gamma_E}c_{\Gamma}}{(16\pi^2)^2}\frac{1}{2\epsilon^4}\frac{1}{t}
\left\{\left(\!\frac{\mu^2}{s}\!\right)^{2\epsilon}\!\!\!\!F_{21}(1,-\epsilon,1\!-\!\epsilon,-\frac ut)\!+\!\left(\!\frac{\mu^2}{t}\!\right)^{2\epsilon}\!\!\!\!\left(\!-1\!+\!F_{21}(\epsilon,2\epsilon,1\!+\!\epsilon,-\frac{s\!+\!u}{t})\!\right)\right.\nonumber \\
&-&\left.
\left(\frac{\mu^2}{s+t+u}\right)^{2\epsilon}\frac{\epsilon}{1+\epsilon}\frac{u}{t+u}
F_1(1,2\epsilon,1,2+\epsilon,\frac{s+u}{s+t+u},\frac{u}{t+u})\right\} \label{G7}
\end{eqnarray}
\beq
_2F_{1}(1,-\epsilon,1-\epsilon,x)=\sum_{n=0}^4\epsilon^n~b_{n}(x)+O(\epsilon^5),
\eeq where
$$
b_0(x)=1
$$
$$
b_1(x)=\log(1-x)
$$
$$
b_2(x)=-Li_2(x)
$$
$$
b_3(x)=-Li_3(x)
$$
$$
b_4(x)=-Li_4(x).
$$
and
\beq
 _2F_{1}(\epsilon,2\epsilon,1+\epsilon,x)=\sum_{n=0}^4\epsilon^n~c_{n}(x)+O(\epsilon^5),\eeq
where
$$
c_0(x)=1
$$
$$
c_1(x)=0
$$
$$
c_2(x)=+2Li_2(x)
$$
$$
c_3(x)=(\frac{2\pi^2}{3}\log(1-x)-2\log^2(1-x)\log(x)-4\log(1-x)Li_2(x)
-4Li_3(1-x)-2Li_3(x)+4\zeta_3),
$$
$$
c_4(x)=\frac{1}{90}(4\pi^4-90\pi^2\log(1-x)^2-15\log(1-x)^4+300\log^3(1-x)\log(x)+360\log^2(1-x)Li_2(x)
$$
$$
+\log(1-x)\textbf{(}-360\zeta_3+720Li_3(1-x)+360Li_3(x)\textbf{)}-360Li_4(1-x)-180Li_4(x)-360Li_4(\frac{x}{x-1})).
$$
In the integral $G_7$ there is the Appell function of the first kind
defined by the  following integral representation: \beq
F_1(\alpha,\beta,\beta',\gamma;x,y) \ = \
\frac{\Gamma(\gamma)}{\Gamma(\alpha)\Gamma(\gamma-\alpha)} \int_0^1
du u^{\alpha-1} (1-u)^{\gamma-\alpha-1} (1-ux)^{-\beta}
(1-uy)^{-\beta'},\eeq which in our case gives us the
one-parametric integral \beq F_1(1,2\epsilon,1,2+\epsilon;x,y) \ = \
(1 + \epsilon) \int_0^1 du (1-u)^{\epsilon} (1-ux)^{-2\epsilon}
(1-uy)^{-1}.\eeq

Expanding the integrand over
$\epsilon$ and  then performing the  integration one gets
\beqa && F_1(1,2\epsilon,1,2+\epsilon;x,y) = (1+\epsilon)
\int_0^1 du \left( 1+(\log (1-u)-2 \log (1-u x)) \epsilon \right. \nonumber \\
\nonumber && \left. \makebox[3em]{} +\left(\frac{1}{2} \log
^2(1-u)-2 \log (1-u x) \log (1-u)+2 \log ^2(1-u x)\right)
\epsilon^2 \right) \nonumber \\
\nonumber && \makebox[3em]{} +\frac{1}{6} \left(\log
^3(1-u)-6 \log (1-u x) \log ^2(1-u)+12 \log ^2(1-u x) \log (1-u) \right. \nonumber \\
\nonumber && \left. \makebox[3em]{} -8 \log ^3(1-u x) \epsilon^3 \right) +
O(\epsilon ^4).\eeqa

Up to the second order in $\epsilon$  it is possible to
evaluate the integrals in terms of logarithms and polylogarithms; however,
in higher orders  new functions appear.

Consider, for example,  the  integral \beq \label{1}
\mathcal{I}_1 \ = \ \int_0^1 du \frac{\log^2(1-u)
\log(1-ux)}{1-uy},\eeq where the parameters satisfy the
condition $0<x<y<1$ .
To evaluate this integral we use the integral representation for
 one of the
logarithms and get \beq \label{2} \int_0^1 du
\int_0^1 da \frac{-ux\log^2(1-u)}{(1-uy)(1-uxa)}.\eeq Now taking the
integral over $u$ one has
 \beqa && \mathcal{I}_1 = -\frac 2y \log
\frac{y-x}{y} Li_3 \frac{-y}{1-y} + 2 \int_0^1 \frac{Li_3
\frac{ax}{ax-1}}{a(ax-y)} da.\eeqa
To find the integral \beq \int_0^1 \frac{Li_3
\frac{ax}{ax-1}}{a(ax-y)} da \eeq  it is useful to  introduce a new variable
$$b \ = \ \frac{ax}{ax-1}$$ then
\beq \int_0^1 \frac{Li_3
\frac{ax}{ax-1}}{a(ax-y)} da = - \frac 1y \int_0^{\frac{x}{x-1}} \frac{Li_3
b}{b(1+\frac{1-y}{y}b)} db\eeq
and  using the identity
$$\frac{1}{b(1+\frac{1-y}{y}b)} \ = \ \frac 1b - \frac{1-y}{y(1+\frac{1-y}{y} b)}$$ one  comes  to  the integral
\beq -\frac 1y \left( \int_0^{-\frac{x}{1-x}} \frac{Li_3 b}{b} db -
\frac{1-y}{y} \int_0^{-\frac{x}{1-x}} \frac{Li_3
b}{1+\frac{1-y}{y}b} db \right).\eeq The first integral is
straightforward and for the second integral one can expand the
polylogarithm in power series and get  the answer in terms of the
function \beq Li_{m,n} (x,y) \ = \ \sum_{j>i>0} \frac{y^j}{j^n}
\frac{x^i}{i^m}.\eeq As a result  one gets \beq \int_0^1 \frac{Li_3
\frac{ax}{ax-1}}{a(ax-y)} da = - \frac 1y \left( Li_4 \left(
-\frac{x}{1-x} \right) + Li_{3,1} \left( -\frac{y}{1-y},
\frac{x(1-y)}{y(1-x)}\right) \right).\eeq

Finally putting everything together we obtain
\beqa && \makebox[-2em]{} \mathcal{I}_1 = -\frac 2y \log
\frac{y\!-\!x}{y} Li_3 \frac{-y}{1\!-\!y} \!-\! \frac 2y  \left( Li_4 \left(
-\frac{x}{1\!-\!x} \right) \!+\! Li_{3,1} \left( -\frac{y}{1\!-\!y},
\frac{x(1\!-\!y)}{y(1\!-\!x)}\right) \right) .\eeqa

Another possibility to expand the Appell function is to use the
 Nested Sums computational tool \cite{Moch}  which represents the Appell
function  as some combination of generalized
polylogarithms. The $\epsilon$ expansion for the
Appell function then takes the form \beqa &&
F_1(1,2\epsilon,1,2+\epsilon;x,y) = -\frac{\log (1\!-\!y)}{y}
+\frac{1}{y} \left(-\log ^2(1\!-\!x)+2 \log
(1\!-\!y) \log (1\!-\!x) \right. \\
\nonumber && \left. \makebox[1em]{} -\frac{1}{2} \log ^2(1-y)-\log
(1-y)-2 Li_2(x)-2 Li_2\left(\frac{x-y}{x-1}\right)+Li_2(y)\right)\epsilon
\nonumber \\ && \makebox[1em]{} + \frac{1}{y}\left(-\frac{1}{6} \log
^3(1-y)+\frac{1}{2} (-\log (y)+2 \log (y-x)-1) \log
^2(1-y)-\frac{1}{6} \pi ^2 \log (1-y) \right. \nonumber \\
\nonumber && \left. \makebox[1em]{} +\log ^2(1-x) (\log (x)-\log
(y)+\log (y-x)-1)-2 Li_2(x)-2 Li_2\left(\frac{x-y}{x-1}\right)  \right. \nonumber \\
\nonumber && \left. \makebox[1em]{} +\log (1-x) \left(\log (1-y)
(2-2 \log (y-x))+\frac{1}{3} \left(6 Li_2(x)+6
Li_2\left(1-\frac{x}{y}\right) \right. \right.  \right. \nonumber \\
\nonumber && \left.  \left. \left. \makebox[1em]{}\!  + 6
Li_2\left(\frac{x-y}{x-1}\right)-6 Li_2(y)+\pi
^2\right)\right)+Li_2(y)\!+\!2 Li_3(x)\!-\!2
Li_3\left(1-\frac{x}{y}\right)\!-\!Li_3(1\!-\!y) \right. \nonumber \\
\nonumber && \left.  \makebox[1em]{} \!+\!2
Li_3\left(\frac{x\!-\!y}{x\!-\!1}\right)\!+\!2
Li_3\left(\frac{y\!-\!1}{x\!-\!1}\right)\!+\!2
Li_3\left(\frac{x-y}{(x\!-\!1) y}\right)\!-\!Li_3(y)\!-\!\zeta
(3)\right) \epsilon^2\!+\! \texttt{Fin}\  \epsilon^3 \!+\!
O(\epsilon^4),\eeqa where \beqa && \makebox[-1em]{} \texttt{Fin} =
\frac 1y \left( -2 Li_{1,1,1,1}(1,\frac xy,1,y) + 2 Li_{1,2,1}(\frac
xy,1,y) - 2 Li_{1,2}(\frac xy,y) +
2 Li_{3,1}(\frac xy,y) \right. \\
\nonumber && \left. + S_{0,3}(y)\! +\! Li_3(y) \!- \!S_{0,4}(y)\!- \!H_{2,2}(y)
\!-\!2 Li_{2,1,1}(1,\frac xy,y) \!- \!H_{1,3}(y) \!+\! 2 Li_{1,1,1,1}(\frac xy,1,1,y) \right. \nonumber \\
\nonumber && \left. - H_{1,2,1}(y) - S_{2,2}(y) - 2
Li_{1,1,2}(1,\frac
xy,y) + 2 Li_{1,1,2}(\frac xy,1,y) + 2 Li_{2,2}(\frac xy,y) + H_{1,2}(y) \right. \nonumber \\
\nonumber && \left.  + 2 Li_{1,3}(\frac xy,y) - S_{1,3}(y) + 2
Li_{1,1,1,}(\frac xy,y) + 2 Li_{2,1,1}(\frac xy,1,y) - Li_4(y) - 2
Li_{2,1}(\frac xy,y) \right. \nonumber \\ \nonumber && \left. +
S_{1,2}(y) \!-\! 2 Li_{1,1,1}(\frac xy,1,y) \!+\! 2 Li_{1,1,1,1}(1,1,\frac
xy,y)\! -\! H_{1,1,2}(y)\! -\! 2 Li_{1,2,1}(1,\frac xy,y) \right).\eeqa

Here we use the following definition of the generalized Goncharov
polylogarithms \cite{Goncharov} \beq Li_{m_1,\ldots,m_k}
(x_1,\ldots,x_k) \ = \ \sum_{i_1>i_2>\ldots>i_k>0}
\frac{x_1^{i_1}}{i_1^{m_1}} \ldots \frac{x_k^{i_k}}{i_k^{m_k}}.\eeq
Apart from these functions we have in expansion the so-called
Nielsen  generalized polylogarithms \beq S_{n,p} \ = \
Li_{n+1,1,\ldots,1} (x,\underbrace{1,\ldots,1}_{p-1})\eeq and also
the harmonic polylogarithms \beq H_{m_1,\ldots,m_k} (x) \ = \
Li_{m_1,\ldots,m_k} (x,\underbrace{1,\ldots,1}_{k-1}).\eeq

In the case of $"1+1"$ dimensional kinematics  the expression for
the integral $G_7$ is simplified and the following limiting cases can
be used: \beq
G_7=G_{6}^{"1+1"}+\frac{e^{-2\epsilon\gamma_E}c_{\Gamma}}{(16\pi^2)^2}~\frac{1}{2\epsilon^4}\left\{-2\epsilon^2
J-\frac{1}{t^{2\epsilon}}\right\}, \eeq where $J$ is the integral
\beq J=\int_0^1dxdy \ \frac{ty^{\epsilon -1}}{(tx+sy+ux
y)^{1+2\epsilon}} \eeq
\begin{eqnarray}
J|_{t=0}&=&0, \\
J|_{s=0}&=&-\frac{1}{2\epsilon}\int_0^1 dy \frac{ty^{\epsilon-1}}{(t
+u y )^{1+2\epsilon}}=-\frac{1}{2\epsilon^2}\frac{1}{t^{2\epsilon}}
\ _2F_{1}(\epsilon,
1 + 2\epsilon,1+\epsilon,-\frac ut) \\
J|_{u=0}&=&-\frac{1}{2\epsilon} \left\{\int_0^1 dy
\frac{y^{\epsilon-1}} {(t +s y )^{2\epsilon}} - \int_0^1 dy
\frac{y^{\epsilon-1}} {(sy)^{2\epsilon}}\right\} \nonumber \\
&& \makebox[2em]{} = -\frac{1}{2\epsilon^2} \frac{1}{s^{2\epsilon}}
- \frac{1}{2\epsilon^2} \frac{1}{t^{2\epsilon}} \
_2F_{1}(\epsilon,2\epsilon,1+\epsilon,-\frac st)
\end{eqnarray}
The finite part of $G_{7}$  is given by the following
expression:
\begin{equation}
(G_7)_{fin}=(G_{6})_{fin}+\frac{1}{(16\pi^2)^2}\frac{1}{2t}
\left\{-1+b_4(-\frac ut)+c_4(-\frac{s+u}{t}) -\frac{u}{t+u}
\texttt{Fin}(\frac{s+u}{s+t+u},\frac{u}{t+u})\right\}.
\end{equation}


\begin{thebibliography}{99}

\bibitem{BeisertYangianRev} N.~Beisert, \emph{On Yangian Symmetry in Planar N=4 SYM},
arXiv:1004.5423v2 [hep-th].

\bibitem{BeisertYangianAmpl} T.~Bargheer, N.~Beisert, W.~Galleas, F.~Loebbert,
T.~McLoughlin, \emph{Exacting N=4 Superconformal Symmetry}, JHEP
\textbf{0911} 056 (2009), arXiv:0905.3738v3 [hep-th]; N.~Beisert,
J.~Henn, T.~McLoughlin, J.~Plefka, \emph{One-Loop Superconformal and
Yangian Symmetries of Scattering Amplitudes in N=4 Super
Yang-Mills}, JHEP \textbf{1004} 085 (2010), arXiv:1002.1733v2
[hep-th].

\bibitem{dualKorch} J.~M.~Drummond, G.~P.~Korchemsky and E.~Sokatchev, \emph{Conformal
properties of four-gluon planar amplitudes and Wilson loops}, Nucl.\
Phys.\  B {\bf 795} (2008) 385, arXiv:0707.0243 [hep-th].

\bibitem{Bern1}
Z.~Bern, L.~J.~Dixon, D.~C.~Dunbar and D.~A.~Kosower, \emph{One-Loop
n-Point Gauge Theory Amplitudes, Unitarity and Collinear Limits},
Nucl.\ Phys.\  B {\bf 425} (1994) 217, arXiv:hep-ph/9403226;
\emph{Fusing gauge theory tree amplitudes into loop amplitudes},
Nucl.\ Phys.\  B {\bf 435} (1995) 59, arXiv:hep-ph/9409265.

\bibitem{yangian}
J.~M.~Drummond, J.~M.~Henn and J.~Plefka, \emph{Yangian symmetry of
scattering amplitudes in $\mathcal{N}=4$ super Yang-Mills theory},
arXiv:0902.2987 [hep-th].

\bibitem{minSurface4point}L.~F.~Alday, J.~M.~Maldacena,
\emph{Gluon scattering amplitudes at strong coupling}, JHEP
\textbf{0706} 064 (2007), arXiv:0705.0303 [hep-th].

\bibitem{Alday:2008yw}
L.~F.~Alday and R.~Roiban, \emph{Scattering Amplitudes, Wilson Loops
and the String/Gauge Theory Correspondence}, Phys.\ Rept.\  {\bf
468} (2008) 153, arXiv:0807.1889 [hep-th].

\bibitem{Alday:2010vh}
L.~F.~Alday, J.~Maldacena, A.~Sever and P.~Vieira, \emph{$Y$-system
for Scattering Amplitudes}, arXiv:1002.2459 [hep-th].

\bibitem{Y-review} A.~Kuniba, T.~Nakanishi, J.~Suzuki \emph{T-systems and Y-systems in integrable
systems}, arXiv:1010.1344v1[hep-th].

\bibitem{Maldacena:2010kp}
J.~Maldacena and A.~Zhiboedov, \emph{Form factors at strong coupling
via a $Y$-system}, arXiv:1009.1139 [hep-th].

\bibitem{vanNeerven:1985ja}
W.~L.~van Neerven, \emph{Infrared Behavior Of On-Shell Form-Factors
In a $\mathcal{N}=4$ Supersymmetric Yang-Mills Field Theory},  Z.\
Phys.\  C {\bf 30}, 595 (1986).

\bibitem{Eden:2010zz}
B.~Eden, G.~P.~Korchemsky and E.~Sokatchev, \emph{From correlation
functions to scattering amplitudes}, arXiv:1007.3246 [hep-th];
\emph{More on the duality correlators/amplitudes}, arXiv:1009.2488
[hep-th].

\bibitem{dual}
A.~Brandhuber, P.~Heslop and G.~Travaglini, \emph{MHV Amplitudes in
$\mathcal{N}=4$ Super Yang-Mills and Wilson Loops}, Nucl.\ Phys.\  B
{\bf 794} (2008) 231, arXiv:0707.1153 [hep-th].\\
J.~M.~Drummond, J.~Henn, G.~P.~Korchemsky and E.~Sokatchev, \emph{On
planar gluon amplitudes/Wilson loops duality}, Nucl.\ Phys.\  B {\bf
795} (2008) 52, arXiv:0709.2368 [hep-th].

\bibitem{Gorsky:2009ew}
A.~Gorsky, \emph{Amplitudes in the $\mathcal{N}=4$ SYM from Quantum
Geometry of the Momentum Space}, Phys.\ Rev.\  D {\bf 80} (2009)
125002, arXiv:0905.2058 [hep-th].

\bibitem{colK} S.~V.~Ivanov, G.~P.~Korchemsky and A.~V.~Radyushkin,
\emph{Infrared Asymptotics Of Perturbative QCD: Contour Gauges},
Yad.\ Fiz.\  {\bf 44} (1986) 230, [Sov.\ J.\ Nucl.\ Phys.\  {\bf 44}
(1986) 145]. \\ G.~P.~Korchemsky and A.~V.~Radyushkin, \emph{Loop
Space Formalism And Renormalization Group For The Infrared
Asymptotics Of QCD}, Phys.\ Lett.\  B {\bf 171} (1986) 459;
\emph{Renormalization of the Wilson Loops Beyond the Leading Order},
Nucl.\ Phys.\  B {\bf 283} (1987) 342.

\bibitem{Transc}
B.~Eden and M.~Staudacher, \emph{Integrability and
transcendentality}, J.\ Stat.\ Mech.\ {\bf 0611} (2006) P014, arXiv:hep-th/0603157.\\
N.~Beisert, B.~Eden and M.~Staudacher, \emph{Transcendentality and
crossing}, J.\ Stat.\ Mech.\  {\bf 0701} (2007) P021,
arXiv:hep-th/0610251.

\bibitem{Kazakov:1987jk}
D.~I.~Kazakov and A.~V.~Kotikov, \emph{Total $\alpha_s$ correction
to deep inelastic scattering cross-section ratio, $R =
\frac{\sigma_L}{ \sigma_t}$ in QCD. Calculation of longitudial
structure function}, Nucl.\ Phys.\ B {\bf 307} (1988) 721
[Erratum-ibid.\  B {\bf 345} (1990) 299].

\bibitem{Kotikov:2003fb}
A.~V.~Kotikov, L.~N.~Lipatov and V.~N.~Velizhanin, \emph{Anomalous
dimensions of Wilson operators in $\mathcal{N} = 4$ SYM theory},
Phys.\ Lett.\ B {\bf 557} (2003) 114, arXiv:hep-ph/0301021; \\
A.~V.~Kotikov, L.~N.~Lipatov, A.~I.~Onishchenko and
V.~N.~Velizhanin, \emph{Three-loop universal anomalous dimension of
the Wilson operators in $\mathcal{N} =  4$ SUSY Yang-Mills model},
Phys.\ Lett.\ B {\bf 595} (2004) 521 [Erratum-ibid.\  B {\bf 632}
(2006) 754], arXiv:hep-th/0404092.

\bibitem{4loopKonsihi}
F.~Fiamberti, A.~Santambrogio, C.~Sieg and D.~Zanon, \emph{Wrapping
at four loops in $\mathcal{N}=4$ SYM}, Phys.\ Lett.\  B {\bf 666}
(2008) 100, arXiv:0712.3522 [hep-th]; \emph{Anomalous dimension with
wrapping at four loops in $\mathcal{N}=4$ SYM}, Nucl.\ Phys.\ B {\bf
805} (2008) 231, arXiv:0806.2095 [hep-th]. \\   Z.~Bajnok and
R.~A.~Janik, \emph{Four-loop perturbative Konishi from strings and
finite size effects for multiparticle states}, Nucl.\ Phys.\  B
{\bf 807} (2009) 625, arXiv:0807.0399 [hep-th].\\
V.~N.~Velizhanin, \emph{The Four-Loop Konishi in $\mathcal{N}=4$
SYM}, arXiv:0808.3832 [hep-th].

\bibitem{N=4onshellmethods} J.~M.~Drummond, J.~Henn,  G.~P.~Korchemsky,
E.~Sokatchev,\emph{Generalized unitarity for $\mathcal{N}=4$
super-amplitudes}, arXiv:0808.0491 [hep-th]; M.~Bianchi, H.~Elvang,
D.~Z.~Freedman, \emph{Generating Tree Amplitudes in N=4 SYM and N =
8 SG}, JHEP 0809:063 2008, arXiv:0805.0757 [hep-th].

\bibitem{He:2010ju}
S.~He and T.~McLoughlin, \emph{On All-loop Integrands of Scattering
Amplitudes in Planar $\mathcal{N}=4$ SYM}, arXiv:1010.6256 [hep-th].

\bibitem{finite1}   L.~V.~Bork, D.~I.~Kazakov, G.~S.~Vartanov and A.~V.~Zhiboedov,
\emph{Infrared Safe Observables in $\mathcal{N}=4$ Super Yang-Mills
Theory}, Phys.\ Lett.\  B {\bf 681} (2009) 296, [arXiv:0908.0387
[hep-th]]; \emph{Construction of Infrared Finite Observables in
$\mathcal{N}=4$ Super Yang-Mills Theory}, Phys.\ Rev.\  D {\bf 81}
(2010) 105028, arXiv:0911.1617 [hep-th].

\bibitem{finite2} L.~V.~Bork, D.~I.~Kazakov, G.~S.~Vartanov and A.~V.~Zhiboedov,
\emph{Infrared Finite Observables in $\mathcal{N}=8$ Supergravity},
Proceedings of the Steklov Institute of Mathematics, to appear {\bf
273} (2011), arXiv:1008.2302 [hep-th].

\bibitem{hofmal} D.~M.~Hofman and J.~Maldacena,
\emph{Conformal collider physics: Energy and charge correlations},
JHEP {\bf 0805} (2008) 012, arXiv:0803.1467 [hep-th].

\bibitem{Magnea:1990zb}  L.~Magnea and G.~F.~Sterman,
\emph{Analytic continuation of the Sudakov form-factor in QCD}, Phys.\ Rev.\  D {\bf 42}, 4222 (1990);\\
L.~J.~Dixon, L.~Magnea and G.~F.~Sterman,
\emph{Universal structure of subleading infrared poles in gauge theory amplitudes},
JHEP {\bf 0808} (2008) 022, arXiv:0805.3515 [hep-ph].

\bibitem{Korchemskaya:1992je}
I.~A.~Korchemskaya and G.~P.~Korchemsky, \emph{On lightlike Wilson
loops}, Phys.\ Lett.\  B {\bf 287}, 169 (1992). \\ A.~Bassetto,
I.~A.~Korchemskaya, G.~P.~Korchemsky and G.~Nardelli, \emph{Gauge
invariance and anomalous dimensions of a light cone Wilson loop in
lightlike axial gauge}, Nucl.\ Phys.\  B {\bf 408}, 62 (1993),
arXiv:hep-ph/9303314.


\bibitem{DelDuca:2009au}
V.~Del Duca, C.~Duhr and V.~A.~Smirnov, \emph{An Analytic Result for
the Two-Loop Hexagon Wilson Loop in $\mathcal{N} = 4$ SYM}, JHEP
{\bf 1003} (2010) 099, arXiv:0911.5332 [hep-ph]; \emph{The Two-Loop
Hexagon Wilson Loop in $\mathcal{N} = 4$ SYM}, JHEP {\bf 1005}
(2010) 084 [arXiv:1003.1702 [hep-th]]; \emph{A Two-Loop Octagon
Wilson Loop in $\mathcal{N} = 4$ SYM}, JHEP {\bf 1009} (2010) 015,
arXiv:1006.4127 [hep-th].

\bibitem{Goncharov:2010jf}
A.~B.~Goncharov, M.~Spradlin, C.~Vergu and A.~Volovich,
\emph{Classical Polylogarithms for Amplitudes and Wilson Loops},
arXiv:1006.5703 [hep-th].

\bibitem{vanNeerven:1985xr}
W.~L.~van Neerven, \emph{Dimensional Regularization Of Mass And
Infrared Singularities In Two Loop On-Shell Vertex Functions},
Nucl.\ Phys.\ B {\bf 268} (1986) 453.

\bibitem{Goncharov} A. B. Goncharov, \emph{Multiple Polylogarithms, cyclotomy and modular
complexes}, Math. Res. Lett. {\bf 5}, 497 (1998); \emph{A simple
construction of Grassmannian polylogarithms}, to appear in the
special volume dedicated to A.Suslin's 60th birthday,     arXiv:0908.2238v2.\\
E.~Remiddi and J.~A.~M.~Vermaseren, \emph{Harmonic polylogarithms},
Int.\ J.\ Mod.\ Phys.\  A {\bf 15} (2000) 725, arXiv:hep-ph/9905237.

\bibitem{Usyukina:1992jd}
N.~I.~Usyukina and A.~I.~Davydychev, \emph{An Approach to the
evaluation of three and four point ladder diagrams}, Phys.\ Lett.\ B
{\bf 298} (1993) 363; \emph{Exact results for three and four point
ladder diagrams with an arbitrary number of rungs}, Phys.\ Lett.\  B
{\bf 305} (1993) 136.

\bibitem{Drummond:2006rz}
J.~M.~Drummond, J.~Henn, V.~A.~Smirnov and E.~Sokatchev, \emph{Magic
identities for conformal four-point integrals}, JHEP {\bf 0701}
(2007) 064, arXiv:hep-th/0607160.

\bibitem{KotTransc} A.~V.~Kotikov and L.~N.~Lipatov, \emph{On the highest
transcendentality in $\mathcal{N}=4$ SUSY}, Nucl.\ Phys.\  B {\bf
769} (2007) 217, arXiv:hep-th/0611204.

\bibitem{1001}S.~J.~Gates, M.~T.~Grisaru, M.~Rocek, W.~Siegel, \emph{Superspace,
or One thousand and one lessons in supersymmetry}, Front.Phys.
\textbf{58} (1983), arXiv:hep-th/0108200v1.

\bibitem{SiegSuperspacenew}C.~Sieg \emph{Superspace computation of the three-loop dilatation
operator of $\mathcal{N}=4$ SYM theory}, arXiv:arXiv:1008.3351
[hep-th].

\bibitem{Penati:1999ba}
S.~Penati, A.~Santambrogio and D.~Zanon, \emph{Two-point functions
of chiral operators in $\mathcal{N} = 4$ SYM at order $g^4$}, JHEP
{\bf 9912} (1999) 006, arXiv:hep-th/9910197; \emph{More on
correlators and contact terms in $\mathcal{N} = 4$ SYM at order
$g^4$},  Nucl.\ Phys.\  B {\bf 593} (2001) 651,
arXiv:hep-th/0005223.

\bibitem{N=4Realesys} S.~Kovacs, \emph{A Perturbative reanalysis of N=4 supersymmetric
Yang-Mills theory}, Int.\ J.\ Mod.\ Phys.\ \textbf{A}21 4598 (2006),
arXiv:hep-th/9902047.

\bibitem{FeynCalc} http://www.feyncalc.org/

\bibitem{Susymath} A. F. Ferrari, \emph{SusyMath: a Mathematica package for quantum
superfield calculations}, Comp. Phys. Comm., 176, 334 (2006);
http://fma.if.usp.br/~alysson/SusyMath

\bibitem{Gehrmann:1999as}
T.~Gehrmann and E.~Remiddi, \emph{Differential equations for
two-loop four-point functions}, Nucl.\ Phys.\  B {\bf 580} (2000)
485, arXiv:hep-ph/9912329; \emph{Two-Loop Master Integrals for
$\gamma^* \to 3$ Jets: The planar topologies}, Nucl.\ Phys.\  B {\bf
601} (2001) 248, arXiv:hep-ph/0008287.

\bibitem{GehrmannCross}
T.~Gehrmann, T.~Huber, D.~Maitre, \emph{Two-loop quark and gluon
form-factors in dimensional regularization}, Phys.\ Lett.\ B {\bf622
} (2005) 295-302, arXiv:hep-ph/0507061.

\bibitem{Anastasiou:1999ui}
C.~Anastasiou, E.~W.~N.~Glover and C.~Oleari, \emph{Scalar One-Loop
Integrals using the Negative-Dimension Approach}, Nucl.\ Phys.\  B
{\bf 572} (2000) 307, arXiv:hep-ph/9907494.

\bibitem{Ellis:2007qk}
R.~K.~Ellis and G.~Zanderighi, \emph{Scalar one-loop integrals for
QCD}, JHEP {\bf 0802} (2008) 002, arXiv:0712.1851 [hep-ph].

\bibitem{Moch}
S. Weinzierl, \emph{Symbolic Expansion of Transcendental Functions},
Comp. Phys. Comm. 145, (2002), 357, math-ph/0201011;\\
S.~Moch, P.~Uwer and S.~Weinzierl, \emph{Nested sums, expansion of
transcendental functions and multi-scale multi-loop integrals}, J.\
Math.\ Phys.\  {\bf 43} (2002) 3363, arXiv:hep-ph/0110083.

\bibitem{Huber:2005yg} T.~Huber and D.~Maitre, \emph{HypExp,
a Mathematica package for expanding hypergeometric functions around
integer-valued parameters}, Comput.\ Phys.\ Commun.\  {\bf 175}
(2006) 122, arXiv:hep-ph/0507094; \emph{HypExp 2, Expanding
Hypergeometric Functions about Half-Integer Parameters}, Comput.\
Phys.\ Commun.\  {\bf 178} (2008) 755, arXiv:0708.2443 [hep-ph].

\bibitem{bds05} Z.~Bern, L.~J.~Dixon and V.~A.~Smirnov,
\emph{Iteration of planar amplitudes in maximally supersymmetric
Yang-Mills theory at three loops and beyond}, Phys.\ Rev.\  D {\bf
72} (2005) 085001, arXiv:hep-th/0505205.

\bibitem{Brandhuber:2010ad}
A.~Brandhuber, B.~Spence, G.~Travaglini and G.~Yang,
\emph{Form Factors in $\mathcal{N}=4$ Super Yang-Mills and Periodic Wilson Loops},
arXiv:1011.1899 [hep-th].

\end{thebibliography}
\end{document}